\documentclass[prd,aps,nofootinbib,floatfix,10 pt]{revtex4}

\usepackage{amssymb}
\usepackage{amsmath,graphicx,color,epsfig,graphics}
\usepackage{slashed}
\usepackage{multirow}
\begin{document}

\title{Effects of neutral $Z^{\prime}$ boson in $B_{s} \to \phi \ell^{+} \ell^{-}$ decay with polarized $\phi$
and the unpolarized and polarized $CP$ violation asymmetry}
\author{Ishtiaq Ahmed$^{1,2}$}
\email{ishtiaq@ncp.edu.pk}
\author{M. Jamil Aslam$^{2}$}
\email{jamil@phys.qau.edu.pk}
\author{M. Ali Paracha$^{3,4}$}
\email{ali@ncp.edu.pk}
\affiliation{$^{1}$National Centre for Physics,\\
Quaid-i-Azam University Campus, Islamabad 45320, Pakistan}
\affiliation{$^{2}$Department of Physics,\\
Quaid-i-Azam University, Islamabad 45320, Pakistan}
\affiliation{$^{3}$Centre for Advanced Mathematics and Physics, \\
National University of Science and Technology, Islamabad, Pakistan}
\affiliation{$^{4}$Laboratorio de F\'{\i}sica Te\'{o}rica e Computacional, Universidade Cruzeiro do Sul,
01506-000, S\H{a}o Paulo, Brazil}

\date{\today}
\begin{abstract}

The effects of new neutral $Z^\prime$ boson in $B_{s}\to\phi\ell^{+}\ell^{-}$, when $\phi$ is longitudinal
or transverse polarized, are studied. In addition, the implications of $Z^{\prime}$ boson on the
unpolarized and polarized $CP$ violation asymmetries, with
reference to lepton, are also presented. It is observed that the branching ratio with polarized $\phi$ are
quite sensitive to the $Z^{\prime}$ contributions which are coming through the modification of Wilson coefficients
$C^{eff}_{9}$ and $C_{10}$. Moreover, the off-diagonal elements of the chiral
$Z^{\prime}$ couplings contain new weak phase $\phi_{sb}$ that provide a new source of $CP$ violation. Keeping in view that in the FCNC
transitions the $CP$ violation asymmetries are highly suppressed in the SM,
we have studied the unpolarized and polarized $CP$ violation asymmetries in
$B_{s}\to\phi\ell^{+}\ell^{-}$ decays. Our results indicate that these $CP$ violation
asymmetries are remarkably significant and can attribute the any new physics coming
through the $Z^{\prime}$ boson. It is hoped that the accurate measurements of these asymmetries will not only
help us to establish NP but will also give a chance to determine the precise values of the coupling parameters of
the $Z^{\prime}$ boson.
\end{abstract}

\maketitle

\section{Introduction}\label{intro}

The purpose of high energy experiments is to resolve the unanswered questions
in the Standard Model (SM) through searches of new physics (NP) using complimentary approaches.
The first approach is at the energy frontiers where the key representatives are
the ATLAS and CMS experiments at the Large Hadron Collider (LHC) at CERN. The main purpose of these two detectors is to
smash the particles at sufficiently large energy and then study the different particles produced after
this collision. The second approach is at the rare/precision frontier where LHCb experiment at the LHC
and the Belle II at the super KEKB are the names of two important experiments in regard to the flavor physics.

In the precision approach the observable signature of new particles or processes can be obtained through
the measurements of flavor physics reactions at lower energies and collect the evidence of any deviation from
these predictions. A natural place is to investigate the flavor-changing-neutral-current (FCNC) processes in
$B$-meson decays where one of the heavy quark makes them an ideal laboratory to test the non-perturbative aspects
of the QCD and also make them a fertile hunting ground for testing the SM and probing the possible NP effects.

Undoubtedly, the predictions of the SM are in good agreement with the collider data until now, however,
still there exists some mysteries that are unanswered in this model. Just to name the few, they include neutrino oscillations,
baryon asymmetry, dark matter, unification, the strong $CP$ violation and the hierarchy problems.
To answer these issues there exists a plethora of the NP models such as the extra dimension models, various
supersymmetric models, etc. In grand unification theories such as $SU(5)$ or string-inspired $E6$ models \cite{S1,S2,S3,S4,S5}, one of the
most pertinent is the $Z^{\prime}$ scenarios that include the family non-universal $Z^{\prime}$ \cite{ZpJ1,ZpJ2} and leptophobic
$Z^{\prime}$ models \cite{ZpJ3,ZpJ4}.

It is well known that the gauge group $SU(5)$ can be extended to the next important group
$SO(10)$ which has one extra rank and hence lead to an idea of an extra heavy neutral $Z$ boson \cite{ZpJ5}.
Even though $Z^{\prime}$ gauge couplings are family universal \cite{ZpJ6,ZpJ7,ZpJ8,ZpJ9,ZpJ10}, however,
due to different constructions of the different
families in string models it is possible to have family non-universal $Z^{\prime}$ couplings.
For example in some
of them, three generation of leptons and also the first and second generation of quarks have
different coupling to $Z^{\prime}$ boson when compared to the third families of quarks \cite{ZpJ2,ZpJ11,ZpJ12}.
The details about this model can be seen for instance in \cite{ZpJ1,ZpJ13,ZpJ14,ZpJ15,ZpJ16,ZpJ17,ZpJ18}.

Searching of an extra $Z^{\prime}$ boson is an important mission of the Tevatron \cite{ZpJ19} and
the LHC \cite{ZpJ20} experiments. Performing constraints on the new $Z^{\prime}$ couplings through low-energy
precise processes is, on the other hand, very crucial and complementary for these direct searches $Z^{\prime} \to e^{+}e^{-}$
at the Tevatron \cite{ZpJ21}. It is interesting to note that such a family non-universal $Z^{\prime}$
model could bring new $CP$-violating phases beyond the SM and have a large effect on many FCNC processes \cite{ZpJ23,ZpJ24},
such as the $B_{s}-\bar{B}_{s}$ mixing \cite{ZpJ25,ZpJ26,ZpJ27,ZpJ28,ZpJ29}, as well as some rare \cite{ZpJ30} and hadronic $B-$meson decays
\cite{ZpJ31, ZpJ31a}.

In the present study we will analyze the $B_{s} \to \phi \ell^{+}\ell^{-}$ decay in the family non-universal
$Z^{\prime}$ scenario. At quark level this decay is governed by the FCNC transition $b\to s \ell^{+}\ell^{-}$, which aries at
loop level in the SM because of the Glashow-Ilipoulos-Maiani (GIM) mechanism where the new heavily
predicted particles of different models can manifest themselves. In particular, by analyzing
 the different physical observables like the decay rate, the forward-backward asymmetry and
different lepton polarization asymmetries and comparing them with the SM predictions one can test the SM as well as
find the traces of the physics beyond it. A detailed analysis of the above mentioned physical observables in
family non-universal $Z^{\prime}$ model for
$B_{s}\to \phi \ell^{+}\ell^{-}$ is discussed in length in ref. \cite{ZpJ32}. However, the study of the polarized and
unpolarized $CP$ violation asymmetries as well as the polarized branching ratio is still missing in the literature.

With the motivation that the behavior of the other observables in the presence of the $Z^{\prime}$ boson may play a crucial
role in redefining our knowledge about the family nonuniversal $Z^{\prime}$ model, we have studied both
polarized and unpolarized $CP$ violation asymmetries and the polarized branching ratio for $B_{s}\to \phi \ell^{+}\ell^{-}$,
in the SM and $Z^{\prime}$ model. In the context of $CP$ violation asymmetry, it is important to emphasize that the
FCNC transitions are proportional to three CKM matrix
elements, namely, $V_{tb}V^{*}_{ts}$, $V_{cb}V^{*}_{cs}$, and
$V_{ub}V^{*}_{us}$; however,
due to the unitarity condition, and neglecting $V_{ub}V^{*}_{us}$ in
comparison of $V_{tb}V^{*}_{ts}$ and $V_{cb}V^{*}_{cs}$, the $CP$ violation asymmetry is
highly suppressed in the SM. Therefore, the measurement
of $CP$ violation asymmetries in FCNC decays play a pivotal role to find the signatures of the
$Z^{\prime}$ model.

This paper is schemed as follows: In section II we briefly describe the
theoretical formulation necessary for the transition $b\to s$,
including, effective Hamiltonian, matrix elements in terms of form
factors and then define the amplitude by using these matrix elements.
In section III we give the explicit expression of polarized branching ratio, polarized and unpolarized $CP$ violation
asymmetries for $B_{s}\to \phi \ell^{+}\ell^{-}$.
Section IV presents the phenomenological analysis and
discussion on the numerical results. The summary of the results and concluding remarks will also be given in the same section.

\section{The $B_{s} \to \phi \ell^{+} \ell^{-}$ transition in the SM and family non-universal $Z^{\prime}$ model}
\subsection{The SM effective Hamiltonian}
The effective Hamiltonian for the decay channel $B_{s} \to \phi \ell^{+} \ell^{-}$ with $\ell = \mu, \tau$, proceed through the quark
level transition $b \to s \ell^{+} \ell^{-}$ in the SM, can be written as
\begin{equation}
H_{eff}=-\frac{4G_{F}}{\sqrt{2}}V_{tb}^{\ast }V_{ts}{\sum\limits_{i=1}^{10}}%
C_{i}({\mu })O_{i}({\mu }),  \label{effective hamiltonian 1}
\end{equation}%
where $G_{F}$ is a Fermi coupling constant and $V_{ij}$ are the matrix elements of
Cabibbo-Kobayashi-Maskawa (CKM) matrix. In above equation (\ref{effective hamiltonian 1})
$O_{i}({\mu })$ $(i=1,\ldots ,10)$ are the four-quark operators and $%
C_{i}({\mu })$ are the corresponding Wilson\ coefficients at the
energy scale ${\mu }$ and the explicit expressions of these Wilson
Coefficients at next-to-leading order (NLO) and
next-to-next-leading logarithm (NNLL) are given in \cite{Buchalla, gfunctions1,Buras1, LCSR, Kruger, Grinstein, Cella, Bobeth, Asatrian, Misiak, Huber}. By considering the fact that $\frac{V_{ub}V^{\ast}_{us}}{V_{tb}V^{\ast}_{ts}}<0.02$, we have neglected the terms proportional to $V_{ub}V^{\ast}_{us}$. The operators responsible for $%
B_{s}\rightarrow \phi\ell^{+}\ell^{-}$ are $O_{7}$, $O_{9}$ and
$O_{10}$ and their form is given by
\begin{eqnarray}
O_{7} &=&\frac{e^{2}}{16\pi ^{2}}m_{b}\left( \bar{s}\sigma _{\mu \nu
}P_{R}b\right) F^{\mu \nu },\,  \notag \\
O_{9} &=&\frac{e^{2}}{16\pi ^{2}}(\bar{s}\gamma _{\mu
}P_{L}b)(\bar{l}\gamma
^{\mu }l),\,  \label{op-form} \\
O_{10} &=&\frac{e^{2}}{16\pi ^{2}}(\bar{s}\gamma _{\mu }P_{L}b)(\bar{l}%
\gamma ^{\mu }\gamma _{5}l),  \notag
\end{eqnarray}%
with $P_{L,R}=\left( 1\pm \gamma _{5}\right) /2$. Neglecting the strange quark mass,
the effective Hamiltonian (\ref{effective hamiltonian 1}) gives the following matrix element
\begin{eqnarray}
\mathcal{M}(B_{s}\rightarrow \phi \ell^{+}\ell^{-})&=&\frac{\alpha_{em} G_F}{2\sqrt{2}\pi}V_{tb}V_{ts}^{\ast}
\bigg[\langle \phi(k,\varepsilon)|\overline{s}\gamma^{\mu}(1-\gamma^5)b|B_{s}(p)\rangle
\left\{C_9^{eff}(\overline{l}\gamma^{\mu}l)+C_{10}(\overline{l}\gamma^{\mu}\gamma^5l)\right\}\notag \\
&&-2C_7^{eff} m_b\langle
\phi(k,\varepsilon)|\overline{s}i\sigma_{\mu\nu}\frac{q^{\nu}}{s}(1+\gamma^5)b|B_{s}(p)\rangle(\overline{l}\gamma^{\mu}l)\bigg],\label{Amplitude}
\end{eqnarray}
where $\alpha_{em}$ is the electromagnetic coupling constant calculated at the
$Z$ boson mass scale. Also, $q=p_1+p_2$ is the momentum transfer to the final lepton pair, with
$p_1$ and $p_2$ are the momentum of $\ell^-$ and
$\ell^+$, respectively and $s$ is the square of the momentum transfer.

The\ Wilson coefficient $C_{9}^{SM}(\mu )$ \ with commonly used notation $%
C_{9}^{eff}(\mu )$ corresponds to the semileptonic operator $O_{9}$. It can
be decomposed into three parts
\begin{equation}
C_{9}^{SM}=C_{9}^{eff}(\mu )=C_{9}(\mu )+Y_{SD}(z,s^{\prime
})+Y_{LD}(z,s^{\prime }),  \label{c9coefficient}
\end{equation}%
where the parameters $z$ and $s^{\prime }$ are defined as $%
z=m_{c}/m_{b},\,\,\,s^{\prime }=q^{2}/m_{b}^{2}$. The
function $Y_{SD}(z,s^{\prime })$ corresponding to short distance describes the perturbative part which
include the indirect contributions from the matrix element of four-quark
operators $\sum_{i=1}^{6}\langle l^{+}l^{-}s|O_{i}|b\rangle $ and this lies
sufficiently far away from the $c\bar{c}$ resonance regions. The manifest
expressions for $Y_{SD}(z,s^{\prime })$ can be written as \cite{53, 54}
\begin{eqnarray}
Y_{SD}(z,s^{\prime }) &=&h(z,s^{\prime })(3C_{1}(\mu )+C_{2}(\mu
)+3C_{3}(\mu )+C_{4}(\mu )+3C_{5}(\mu )+C_{6}(\mu ))  \notag \\
&&-\frac{1}{2}h(1,s^{\prime })(4C_{3}(\mu )+4C_{4}(\mu )+3C_{5}(\mu
)+C_{6}(\mu ))  \notag \\
&&-\frac{1}{2}h(0,s^{\prime })(C_{3}(\mu )+3C_{4}(\mu ))+{\frac{2}{9}}%
(3C_{3}(\mu )+C_{4}(\mu )+3C_{5}(\mu )+C_{6}(\mu )),  \label{short-distance}
\end{eqnarray}%
with
\begin{eqnarray}
h(z,s^{\prime }) &=&-{\frac{8}{9}}\mathrm{ln}z+{\frac{8}{27}}+{\frac{4}{9}}x-%
{\frac{2}{9}}(2+x)|1-x|^{1/2}\left\{
\begin{array}{l}
\ln \left\vert \frac{\sqrt{1-x}+1}{\sqrt{1-x}-1}\right\vert -i\pi \quad
\mathrm{for}{{\ }x\equiv 4z^{2}/s^{\prime }<1} \\
2\arctan \frac{1}{\sqrt{x-1}}\qquad \mathrm{for}{{\ }x\equiv
4z^{2}/s^{\prime }>1}%
\end{array}%
\right. ,  \notag \\
h(0,s^{\prime }) &=&{\frac{8}{27}}-{\frac{8}{9}}\mathrm{ln}{\frac{m_{b}}{\mu
}}-{\frac{4}{9}}\mathrm{ln}s^{\prime }+{\frac{4}{9}}i\pi \,\,.  \label{hzs}
\end{eqnarray}%
The long-distance contributions $Y_{LD}(z,s^{\prime })$ from four-quark
operators near the $c\bar{c}$ resonance cannot be calculated from first
principles of QCD and are usually parameterized in the form of a
phenomenological Breit-Wigner formula making use of the vacuum saturation
approximation and quark-hadron duality. In the present study we ignore this
part because this lies far away from the region of interest.

The Wilson coefficient $C_{7}^{eff}$ is given by \cite{68, 69, 70}
\begin{equation}
C_{7}^{SM}=C_{7}^{eff}(\mu )=C_{7}(\mu )+C_{b\rightarrow s\gamma }(\mu ),
\label{c7coefficient}
\end{equation}%
with
\begin{eqnarray}
C_{b\rightarrow s\gamma }(\mu ) &=&i\alpha _{s}\bigg[{\frac{2}{9}}\eta
^{14/23}(G_{1}(x_{t})-0.1687)-0.03C_{2}(\mu )\bigg], \\
G_{1}(x_{t}) &=&{\frac{x_{t}(x_{t}^{2}-5x_{t}-2)}{8(x_{t}-1)^{3}}}+{\frac{%
3x_{t}^{2}\mathrm{ln}^{2}x_{t}}{4(x_{t}-1)^{4}}},
\end{eqnarray}%
where $\eta =\alpha _{s}(m_{W})/\alpha _{s}(\mu )$, $%
x_{t}=m_{t}^{2}/m_{W}^{2}$, $C_{b\rightarrow s\gamma }$ is the absorptive
part for the $b\rightarrow sc\bar{c}\rightarrow s\gamma $ rescattering and
we have dropped out the tiny contributions proportional to CKM sector $%
V_{ub}V_{us}^{\ast }$.

\subsection{The effective Hamiltonian in $Z^{\prime}$ model}
In the $Z^{\prime}$ model, the presence of off-diagonal couplings make the FCNC transitions occur at tree level.
Ignoring the $Z-Z^\prime$
mixing and the interaction of right handed quark with $Z^\prime$ the new gauge boson contribution
only modifies the Wilson coefficients $C_9$ and $C_{10}$ \cite{rhc}. With
these assumptions, the extra part that is added to the Hamiltonian given in Eq. (\ref{effective hamiltonian 1})
can be written as follows \cite{ZPphenomenology,ZPEH1,ZPEH2}
\begin{eqnarray}
\mathcal{H}^{Z^\prime}_{eff}&=&-\frac{2G_F}{\sqrt{2}}\overline{s}\gamma^{\mu}(1-\gamma^5)b \mathcal{B}_{sb}\bigg[\mathcal{S}_{\ell\ell}^L\bar{\ell}\gamma^\mu(1-\gamma^5)\ell-\mathcal{S}_{\ell\ell}^R\bar{\ell}\gamma^\mu(1+\gamma^5)\ell\bigg]+h.c.,\label{Zphamiltonian}
\end{eqnarray}
where $\mathcal{B}_{sb}=|\mathcal{B}_{sb}|e^{-i\phi_{sb}}$ is the off diagonal left handed coupling of
$Z^\prime$ boson with quarks and $\phi_{sb}$ corresponds to a new weak phase.
The left and right handed couplings of the $Z^\prime$ boson with leptons are represented by $\mathcal{S}_{\ell\ell}^L$ and
$\mathcal{S}_{\ell\ell}^R$, respectively.  Therefore, one can
also write the above equation in the following way
\begin{eqnarray}
\mathcal{H}^{Z^\prime}_{eff}&=&-\frac{4G_{F}}{\sqrt{2}}V_{tb}V^\ast_{ts}\bigg[\Lambda_{sb}C_{9}^{Z^\prime}\mathcal{O}_9+\Lambda_{sb}C_{10}^{Z^\prime}\mathcal{O}_{10}\bigg]+h.c.,
\end{eqnarray}
with
\begin{eqnarray}
\Lambda_{sb}&=&\frac{4\pi e^{-i\phi_{sb}}}{\alpha_{s} V_{tb}V_{ts}^\ast},\label{Lfunction}
\\
C_{9}^{Z^\prime}&=&|\mathcal{B}_{sb}|S_{LL};\
C_{10}^{Z^\prime}=|\mathcal{B}_{sb}|D_{LL},\label{C9zdefinition}
\\
S_{LL}&=&\mathcal{S}_{\ell\ell}^L+\mathcal{S}_{\ell\ell}^R;\
D_{LL}=\mathcal{S}_{\ell\ell}^L-\mathcal{S}_{\ell\ell}^R.
\end{eqnarray}
In short to include the $Z^\prime$ effects in the problem under
consideration one has to make the following replacements to the $Z$ boson
Wilson coefficients $C_9$ and $C_{10}$, while, $C_{7}$ remains
unchanged
\begin{eqnarray}
C_{9}^{\prime} &=&C_{9}^{eff} +\Lambda_{sb}C_{9}^{Z^\prime},\notag
\\
C_{10}^{\prime} &=&C_{10} +\Lambda_{sb}C_{10}^{Z^\prime}.
\end{eqnarray}

\subsection{Matrix elements and form factors}
The $B_{s}\rightarrow \phi \ell^{+}\ell^{-}$ decay can be obtained by sandwiching the effective Hamiltonian between
initial state $B_{s}$ and final state $\phi$ meson. This can be parameterized in terms of the form factors as
follows:
\begin{eqnarray}
\langle \phi (k,\varepsilon )|\bar{s}\gamma _{\mu }b|B_{s}(p)\rangle
&=&\varepsilon _{\mu \nu \rho \sigma }\varepsilon ^{\ast \nu }p^{\rho
}k^{\sigma }\frac{2V\left( q^{2}\right) }{M_{B_{s}}+M_{\phi }},\,\,
\label{vectormatrix element} \\
\,\langle \phi (k,\varepsilon )|\bar{s}\gamma _{\mu }\gamma
_{5}b|B_{s}(p)\rangle &=&i\varepsilon _{\mu }^{\ast }\left(
M_{B_{s}}+M_{\phi }\right) A_{1}\left( q^{2}\right) -i(p+k)_{\mu
}(\varepsilon ^{\ast }\cdot q)\frac{A_{2}\left( q^{2}\right) }{%
M_{B_{s}}+M_{\phi }}  \notag \\
&&-iq_{\mu }(\varepsilon ^{\ast }\cdot q)\frac{2M_{\phi }}{q^{2}}\left[
A_{3}\left( q^{2}\right) -A_{0}\left( q^{2}\right) \right] ,\,\,
\label{axialvector-melement} \\
\langle \phi \left( k,\varepsilon \right) |\bar{s}\sigma _{\mu \nu }q^{\nu
}b|B_{s}\left( p\right) \rangle &=&i\varepsilon _{\mu \nu \rho \sigma
}\varepsilon ^{\ast \nu }p^{\rho }k^{\sigma }2T_{1}\left( q^{2}\right) ,\,\,
\label{tensor matrix element 1} \\
\langle \phi \left( k,\varepsilon \right) |\bar{s}\sigma _{\mu \nu }\gamma
_{5}q^{\nu }b|B_{s}\left( p\right) \rangle &=&T_{2}\left( q^{2}\right) \left[
\varepsilon _{\mu }^{\ast }\left( M_{B_{s}}^{2}-M_{\phi }^{2}\right)
-(p+k)_{\mu }(\varepsilon ^{\ast }\cdot q)\right]  \notag \\
&&+T_{3}\left( q^{2}\right) (\varepsilon ^{\ast }\cdot q)\left[ q_{\mu }-%
\frac{q^{2}}{M_{B_{s}}^{2}-M_{\phi }^{2}}(p+k)_{\mu }\right] ,
\label{tensor matrix element 2}
\end{eqnarray}%
where $\varepsilon ^{\ast\nu }$ is the polarization of the final state vector meson $\left( \phi
\right) $.

The form factors $A_{i}$ and $T_{i}$ are functions of the square
of momentum transfer $q$ and these are not independent of each other. By contracting the
above equations with $q_{\mu}$ and making use of the equation of motion, one can write
\begin{eqnarray}
A_{3}\left( q^{2}\right) &=&\frac{M_{B_{s}}+M_{\phi }}{2M_{\phi }}%
A_{1}\left( q^{2}\right) -\frac{M_{B_{s}}-M_{\phi }}{2M_{\phi }}A_{2}\left(
q^{2}\right),  \notag \\
A_{3}\left( 0\right) &=&A_{0}\left( 0\right) \text{, }T_{1}\left( 0\right)
=T_{2}\left( 0\right) .  \label{form-factor-relation}
\end{eqnarray}%
The form factors for $B_{s}\rightarrow \phi $ transition are the
non-perturbative quantities and are the major candidate to prone the uncertainties. In literature there exist different
approaches (both perturbative and non-perturbative) like Lattice QCD, QCD
sum rules, Light Cone sum rules, etc., to calculate them. Here, we will consider the form
factors calculated by using the Light Cone Sum Rules approach by Ball \textit{et al.} \cite{PBALL}. The form factors $V,A_{0}$ and $T_{1}$ are parameterized by
\begin{equation}
F(q^{2})=\frac{r_{1}}{1-q^{2}/m_{R}^{2}}+\frac{r_{2}}{1-q^{2}/m_{fit}^{2}},
\label{form-factor-sq}
\end{equation}%
while the form factors $A_{2}$ and $\tilde{T}_{3}$ are parameterized as
follows,
\begin{equation}
F(q^{2})=\frac{r_{1}}{1-q^{2}/m^{2}}+\frac{r_{2}}{(1-q^{2}/m^{2})^{2}}.
\label{form-factor-sq1}
\end{equation}%
The fit formula for $A_{1}$ and $T_{2}$ is
\begin{equation}
F(q^{2})=\frac{r_{2}}{1-q^{2}/m_{fit}^{2}}.  \label{form-factor-sq2}
\end{equation}%
The form factor $T_{3}$ can be obtained through the relation
\begin{equation*}
T_{3}(q^{2})=\frac{M_{B_{s}}^{2}-M_{\phi }^{2}}{q^{2}}[\tilde{T_{3}}%
(q^{2}-T_{2}(q^{2})],
\end{equation*}%
where the values of different parameters are summarized in Table I.
\begin{table}[tbh]
\caption{Fit parameters for $B_{s}\rightarrow \protect\phi $ transition form
factors. $F(0)$ denotes the value of form factors at $q^{2}=0$ (c.f. (Eq. (\protect
\ref{form-factor-sq})). The theoretical uncertainty estimated
is around $15\%$.}
\label{di-fit B_s to phi)}%
\begin{tabular}{ccccccc}
\hline\hline
& $F(q^{2})$ & $\hspace{1 cm} F(0)$ & $\hspace{1 cm} r_{1}$ & $\hspace{1 cm}
m_{R}^{2}$ & $\hspace{1 cm} r_{2}$ & $\hspace{1 cm} m_{fit}^{2}$ \\ \hline
& $A_{1}(q^{2})$ & $\hspace{1 cm} 0.311$ & $\hspace{1 cm}-$ & $\hspace{1 cm}%
- $ & $\hspace{1 cm} 0.308$ & $\hspace{1 cm} 36.54$ \\ \hline
& $A_{2}(q^{2})$ & $\hspace{1 cm} 0.234$ & $\hspace{1 cm} -0.054$ & $\hspace{%
1 cm} -$ & $\hspace{1 cm} 0.288$ & $\hspace{1 cm} 48.94$ \\ \hline
& $A_{0}(q^{2})$ & $\hspace{1 cm} 0.474$ & $\hspace{1 cm} 3.310$ & $\hspace{%
1 cm} 5.28^{2}$ & $\hspace{1 cm} -2.835$ & $\hspace{1 cm} 31.57$ \\ \hline
& $V(q^{2})$ & $\hspace{1 cm} 0.434$ & $\hspace{1 cm} 1.484$ & $\hspace{1 cm}
5.32^{2}$ & $\hspace{1 cm} -1.049$ & $\hspace{1 cm} 39.52$ \\ \hline
& $T_{1}(q^{2})$ & $\hspace{1 cm} 0.349$ & $\hspace{1 cm} 1.303$ & $\hspace{%
1 cm} 5.32^{2}$ & $\hspace{1 cm} -0.954$ & $\hspace{1 cm} 38.28$ \\ \hline
& $T_{2}(q^{2})$ & $\hspace{1 cm} 0.349$ & $\hspace{1 cm} -$ & $\hspace{1 cm}
-$ & $\hspace{1 cm} 0.349$ & $\hspace{1 cm} 37.21$ \\ \hline
& $\tilde{T}_{3}(q^{2})$ & $\hspace{1 cm} 0.349$ & $\hspace{1 cm} 0.027$ & $%
\hspace{1 cm} -$ & $\hspace{1 cm} 0.321$ & $\hspace{1 cm} 45.56$ \\
\hline\hline
&  &  &  &  &  &
\end{tabular}%
\end{table}

Hence by using the above given matrix elements which are parameterized in terms of
the form factors the decay amplitude for $B_{s}\rightarrow \phi \ell^+\ell^-$ can be written as
\begin{eqnarray}
\mathcal{M}=\frac{\alpha G_F}{4\sqrt{2}\pi}V_{tb} V_{ts}^\ast\bigg[\overline{l}\gamma^\mu(1-\gamma^5)l\times\big(-2\mathcal{J}_{1}\epsilon_{\mu\nu\lambda\sigma}\epsilon^{\ast\nu} k^\lambda q^\sigma-i\mathcal{J}_2\epsilon^\ast_\mu
+i\mathcal{J}_3\epsilon^\ast\cdot q(p+k)_\mu+i\mathcal{J}_4\epsilon^\ast\cdot qq_\mu\big)\notag \\
+\overline{l}\gamma^{\mu}(1+\gamma^5)
l\times\big(-2\mathcal{J}_{5}\epsilon_{\mu\nu\lambda\sigma}\epsilon^{\ast\nu}
k^\lambda
q^\sigma
-i\mathcal{J}_6\epsilon^\ast_\mu+i\mathcal{J}_7\epsilon^\ast\cdot
q(p+k)_\mu+i\mathcal{J}_8\epsilon^\ast\cdot
qq_\mu\big)\bigg].\notag \\
\label{Amplitude2}
\end{eqnarray}
Keeping the final state leptons' mass, we can see that the first line of the above equation will
survive only for $\overline{\ell}\gamma^{\mu}\gamma^5\ell$ due to the fact
that
$q_\mu(\overline{\ell}\gamma^{\mu}\gamma^5\ell)=2m_l(\overline{\ell}\gamma^5\ell)$
and will vanish for $\overline{l}\gamma^{\mu}l$ because of
$q_\mu(\overline{l}\gamma^{\mu}l)=0$.

The auxiliary
functions $\mathcal{J}_{1},\cdot\cdot\cdot,\mathcal{J}_{8}$ contain both long and short distance physics
which is encapsulated in the form factors and in the Wilson
coefficients, respectively. These functions can be written in the following
form:
\begin{eqnarray}
\mathcal{J}_{1}&=&2C_{LL}\mathcal{D}_1+4m_bC_7^{eff}\frac{T_1(s)}{s},\notag \\
\mathcal{J}_2&=&2C_{LL}\mathcal{D}_3+\frac{4m_b}{s}C_7^{eff}\mathcal{D}_4\notag\\
\mathcal{J}_3&=&2C_{LL}\mathcal{D}_6+4\frac{m_bC_7^{eff}}{s}\mathcal{D}_5\notag \\
\mathcal{J}_4&=&\frac{2M_{\phi}}{s}\mathcal{D}_7-\frac{4m_b}{s}C_7^{eff}T_3(s)\notag\\
\mathcal{J}_{5}&=&\mathcal{J}_1(C_{LL}\rightarrow C_{LR})\qquad\notag\\
\mathcal{J}_6&=&\mathcal{J}_2(C_{LL}\rightarrow C_{LR})\notag\\
\mathcal{J}_7&=&\mathcal{J}_3(C_{LL}\rightarrow C_{LR})\qquad\notag\\
\mathcal{J}_8&=&\mathcal{J}_4(C_{LL}\rightarrow
C_{LR}),\label{jfunction}
\end{eqnarray}
where $C_{LL}, C_{LR}$ and $\mathcal{D}$'s are defined as follows
\begin{eqnarray}
 C_{LL}&=&C_9^{\prime}-C_{10}^{\prime},\qquad
 C_{LR}=C_9^{\prime}+C_{10}^{\prime},\notag\\
 \mathcal{D}_1&=&\frac{V(s)}{(M_{B_s}+M_\phi)},\notag\\
 \mathcal{D}_3&=&(M_{B_{s}}+M_{\phi})A_1(s),\notag\\
  \mathcal{D}_4&=&(M_{B_s}^2-M_{\phi}^2)T_2(s),\notag\\
 \mathcal{D}_5&=&\bigg[T_2(s)+\frac{s}{(M_{B_s}^2-M^2_{\phi})}T_3(s)\bigg],\label{D-definition}\\
 \mathcal{D}_6&=&\frac{A_2(s)}{(M_{B_{s}}+M_{\phi})},\notag\\
 \mathcal{D}_7&=&(A_3-A_0).\notag
\end{eqnarray}
Now with all the ingredients in hand the next step is to summarize the formulas of different physical observables.
\section{Formulas of physical observables}
\subsection{Differential Decay Rate}
In order to calculate the polarized branching ration as well as the unpolarized and polarized $CP$ violation asymmetries, we fist have to find the expression for the
differential decay width of $B_{s}\rightarrow \phi \ell^+\ell^-$ decay. The formula for the double differential
decay rate can be written as
\begin{eqnarray}
 \frac{d^2\Gamma(B_{s}\rightarrow \phi \ell^+\ell^-)}{d\cos\theta ds}=\frac{1}{2M_{B_s}^3}\frac{2\beta\sqrt{\lambda}}{(8\pi)^3}|\mathcal{M}|^2,
\end{eqnarray}
where $\beta\equiv\sqrt{1-\frac{4m_l^2}{s}}$ and $\lambda=\lambda(M_{B_s},M_{\phi},s)\equiv
M_{B_s}^4+M_{\phi}^4+s^2-2M_{B_s}^2M_{\phi}^2-2sM_{B_s}^2-2sM_{\phi}^2$. Also $s$ is just square
of the momentum transfer $q$ and $\theta$ is the angle between lepton and final state meson in the rest frame of
$B_{s}$. By using the expression of the decay amplitude given in Eq. (\ref{Amplitude2}) and integrating on the $cos\theta$, one can get the expression of the dilepton invariant mass spectrum as
\begin{eqnarray}
 \frac{d\Gamma(B_{s}\rightarrow \phi \ell^+\ell^-)}{ds}=\frac{G_F^2\alpha^2\beta\sqrt{\lambda}M_{B_s}}{2^{14}\pi^5}|V_{tb}V_{ts}^\ast|^2\mathcal{M}_1\label{decay},
\end{eqnarray}
with
\begin{eqnarray}
\mathcal{M}_1&=&4(2m_l^2+s)\{\frac{8\lambda}{3}\mathcal{R}e|\mathcal{J}_1|^2+\frac{12M_{\phi}^2s+\lambda}{3M_{\phi}^2s}\mathcal{R}e|\mathcal{J}_2|^2
-\frac{(M_{B_s}^2-M_{\phi}^2-s)}{3M_{\phi}^2s}\lambda\mathcal{R}e(\mathcal{J}_2\mathcal{J}^*_3)+\frac{\lambda^2}{3M_{\phi}^2s}\mathcal{R}e|\mathcal{J}_3|^2\}\notag
\\
&&+\frac{32\lambda}{3}(s-4m_l^2)\mathcal{R}e|\mathcal{J}_5|^2+[\frac{4\lambda(2m_l^2+s)}{3M_{\phi}^2s}+16(s-4m_l^2)]\times\mathcal{R}e|\mathcal{J}_6|^2-\frac{4\lambda}{3M_{\phi}^2s}\{[(2m_l^2+s)(M_{B_s}^2-M_{\phi}^2)\notag
\\
&&+s(s-4m_l^2)]\mathcal{R}e(\mathcal{J}_6\mathcal{J}^*_7)+[6m_l^2s(2M_{B_s}^2+2M_{\phi}^2-s)
+\lambda(2m_l^2+s)]\mathcal{R}e|\mathcal{J}_7|^2+\frac{8m_l^2\lambda}{M_{\phi}^2}(M_{B_s}^2-M_{\phi}^2)\mathcal{R}e(\mathcal{J}_7\mathcal{J}^*_8)\notag\\
&&-s\frac{8m_l^2\lambda}{M_{\phi}^2}\mathcal{R}e|\mathcal{J}_8|^2.
\end{eqnarray}
Here we take the liberty to correct the expression of the decay rate given in \cite{ZpJ31a}.
\subsection{Branching ratio of $B_{s}\to\phi\ell^+\ell^-$ with polarized $\phi$}
The total decay rate for $B_{s}\to \phi \ell^+ \ell^-$ can be written in terms of the longitudinal $(\Gamma_L)$
and normal components $(\Gamma_T)$, when the final state vector meson is polarized. The explicit expressions of the differential
decay rate in terms of these components can be written as \cite{phipol}
\begin{equation}
\frac{d\Gamma (s)}{ds}=\frac{d\Gamma _{L}(s)}{ds}+\frac{%
d\Gamma _{T}(s)}{ds},
\end{equation}%
where%
\begin{equation*}
\frac{d\Gamma _{T}(s)}{ds}=\frac{d\Gamma _{+}(s)}{ds}+\frac{%
d\Gamma _{-}(s)}{ds},
\end{equation*}%
and
\begin{eqnarray}
\frac{d\Gamma _{L}(s)}{ds} &=&\frac{G_{F}^{2}\left\vert
V_{tb}V_{ts}^{\ast }\right\vert ^{2}\alpha ^{2}}{2^{11}\pi ^{5}}\frac{%
\beta\sqrt{\lambda}}{M_{B_{s}}^{3}}\times \frac{1}{3}\mathcal{A}_{L},  \label{65d} \\
\frac{d\Gamma _{\pm }(s)}{ds} &=&\frac{G_{F}^{2}\left\vert
V_{tb}V_{ts}^{\ast }\right\vert ^{2}\alpha ^{2}}{2^{11}\pi^{5}}\frac{%
\beta\sqrt{\lambda}}{M_{B_{s}}^{3}}\times \frac{4}{3}\mathcal{A}_{\pm }.  \label{65h}
\end{eqnarray}

The different functions appearing in Eqs. (\ref{65d}) and (\ref{65h}) can be
written as
\begin{align}
\mathcal{A}_{L}& =\frac{1}{sM_{\phi }^{2}}\bigg[24\left\vert
\mathcal{K}_7(s)\right\vert ^{2}m_l^{2}M_{\phi }^{2}\lambda
+(2m_l^{2}+s)\left\vert (M_{B_{s}}^{2}-M_{\phi
}^{2}-s)\mathcal{K}_{2}(s)+\lambda \mathcal{K}_{3}(s)\right\vert ^{2}  \notag \\
& +(s-4m_l^{2})\left\vert (M_{B_{s}}^{2}-M_{\phi
}^{2}-s)\mathcal{K}_{5}(s)+\lambda \mathcal{K}_{6}(s)\right\vert ^{2}\bigg], \label{65j}
\end{align}%
\begin{equation}
\mathcal{A}_{\pm }=(s-4m_l^{2})\left\vert \mathcal{K}_{5}(s)\mp \sqrt{\lambda }%
\mathcal{K}_{4}(s)\right\vert ^{2}+\left( s+2m_l^{2}\right) \left\vert
\mathcal{K}_2(s)\pm \sqrt{\lambda }\mathcal{K}_{1}(s)\right\vert ^{2},  \label{65k}
\end{equation}
with
\begin{align}
\mathcal{K}_1(s)=& 4C_7^{eff}\frac{m_{b}}{s}T_{1}(s)+%
2C_9^{\prime}\mathcal{D}_1, \label{621a} \\
\mathcal{K}_2(s)=& 2\frac{m_{b}}{s}C_7^{eff}\mathcal{D}_4 +C_{9}^{\prime}\mathcal{D}_3, \label{621b} \\
\mathcal{K}_{3}(s)=& 4C_7^{eff}\mathcal{D}_5 +C_9^{\prime}\mathcal{D}_6, \label{621c} \\
\mathcal{K}_{4}(s)=& 2C_{10}^{\prime}\mathcal{D}_1, \label{621d} \\
\mathcal{K}_{5}(s)=& 2C_{10}^{\prime}A_{0}(s)\left( M_{B_{s}}+M_{\phi
}\right),   \label{621e} \\
\mathcal{K}_{6}(s)=& 2C_{10}^{\prime}\frac{\mathcal{D}_3}{(M_{B_{s}}+M_{\phi })^2},
\label{621f} \\
\mathcal{K}_7(s)=& 4C_{10}^{\prime}\mathcal{D}_6 \label{621g}.
\end{align}%
In the above equations the functions $\mathcal{D}_1,\ldots,\mathcal{D}_6$ are given in Eq. (\ref{D-definition}).

\subsection{Polarized and Unpolarized $CP$ violation Asymmetries}

The non-equality of the decay rates of a particle and its antiparticle defines the $CP$ violation
asymmetry. The $CP$ violation asymmetry arises both for the cases when the final state leptons are unpolarized or
polarized. In case of the unpolarized leptons, the normalized $CP$ violation asymmetries can be defined through the
difference of the differential decay rates of particle and
antiparticle decay modes as follows \cite{formcp1,formcp2} \begin{eqnarray}
\mathcal{A}_{CP}(\textbf{S}^\pm=\textbf{e}_i^\pm)=\frac{\frac{d\Gamma(\textbf{S}^-)}{ds}-\frac{d\bar\Gamma(\textbf{S}^+)}{ds}}{\frac{d\Gamma}{ds}-\frac{d\bar\Gamma}{ds}},
\end{eqnarray}
where
\begin{eqnarray}
\frac{d\Gamma(\textbf{S}^-))}{ds}=\frac{d\Gamma(B_{s}\to
\phi\ell^+\ell^-(\textbf{S}^-))}{ds},\notag
\\
\frac{d\bar\Gamma(\textbf{S}^+))}{ds}=\frac{d\bar\Gamma(B_{s}\to
\phi\ell^+(\textbf{S}^+))\ell^-}{ds}.\notag
\end{eqnarray}
The differential decay rate of $B_{s}\to \phi \ell^+\ell^-$ is given in Eq.
(\ref{decay}), analogously the $CP$ conjugated differential decay width can be
written as
\begin{eqnarray}
 \frac{d\bar\Gamma(\textbf{S}^\pm)}{ds}=\frac1 2\left(\frac{d\bar\Gamma}{ds}\right)\bigg[1+(P_L\textbf{e}^\pm_L+P_N\textbf{e}^\pm_N+P_T\textbf{e}^\pm_T)\cdot
 \textbf{S}^\pm\bigg].\notag
\end{eqnarray}
It is noted here, that $\frac{d\bar\Gamma}{ds}$ belongs to the
transition $\bar {B_{s}}\to\bar {\phi}\ell^+\ell^-$ which can be obtained by
replacing $\Lambda_{sb}$ to $\Lambda_{sb}^\ast$ in Eq. (\ref{Lfunction}).
Furthermore, by using the fact that $\textbf{S}^+=-\textbf{S}^-$ for the
longitudinal $(L)$, normal $(N)$ and $\textbf{S}^+=\textbf{S}^-$ for the transverse $(T)$ polarizations, we get
\begin{eqnarray}
\mathcal{A}_{CP}(\textbf{S}^\pm=\textbf{e}_i^\pm)&=&\frac{1}{2}\bigg[\frac{\left(\frac{d\Gamma}{ds}\right)-\left(\frac{d\bar\Gamma}{ds}\right)}{\left(\frac{d\Gamma}{ds}\right)+\left(\frac{d\bar\Gamma}{ds}\right)}%
\pm\frac{\left(\frac{d\Gamma}{ds}\right)P_i-\{\left(\frac{d\Gamma}{ds}\right)P_i\}_{\Lambda_{sb}\to\Lambda_{sb}^\ast}}{\left(\frac{d\Gamma}{ds}\right)+\left(\frac{d\bar\Gamma}{ds}\right)}\bigg],\notag
\end{eqnarray}
where $i$ denotes the $L$, $N$ or $T$ polarizations of the final
state leptons. By using Eq. (\ref{D-definition}) in the above equation, the expression of $CP$ violation
asymmetry becomes
\begin{eqnarray}
\mathcal{A}_{CP}(\textbf{S}^\pm=\textbf{e}_i^\pm)&=&\frac{1}{2}\bigg[\frac{\mathcal{M}_1-\bar{\mathcal{M}_1}}{\mathcal{M}_1+\bar{\mathcal{M}_1}}\pm\frac{\mathcal{M}_1^i-\bar{\mathcal{M}_1^i}}{\mathcal{M}_1^i+\bar{\mathcal{M}_1^i}}\bigg],\label{CP}
\end{eqnarray}
where
$\bar{\mathcal{M}_1}=\left(\mathcal{M}_1\right)_{\Lambda_{sb}\to\Lambda_{sb}^\ast}$,
$\bar{\mathcal{M}_1^i}=\left(\mathcal{M}_1^i\right)_{\Lambda_{sb}\to\Lambda_{sb}^\ast}$
and
\begin{eqnarray}
\mathcal{A}_{CP}(s)=\frac{\mathcal{M}_1-\bar{\mathcal{M}_1}}{\mathcal{M}_1+\bar{\mathcal{M}_1}},\quad\mathcal{A}^i_{CP}(s)=\frac{\mathcal{M}_1^i-\bar{\mathcal{M}_1^i}}{\mathcal{M}_1^i+\bar{\mathcal{M}_1^i}}.
\end{eqnarray}
Hence, by using these definitions the normalized $CP$ violation asymmetry can
be written as follows
\begin{eqnarray}
\mathcal{A}_{CP}(\textbf{S}^\pm=\textbf{e}_i^\pm)&=&\frac{1}{2}\bigg[\mathcal{A}_{CP}(s)\pm\mathcal{A}^i_{CP}(s)\bigg]\label{CPdefined},
\end{eqnarray}
where the plus sign in the second term of the above expression
corresponds to $L$ and $N$ polarizations, and the negative sign
is for the $T$ polarization.

The first term in $\mathcal{A}_{CP}(s)$ in Eq. (\ref{CPdefined}) is the
unpolarized $CP$ violation asymmetry, while the second term
$\mathcal{A}^i_{CP}(s)$ is called the polarized $CP$ violation
asymmetry which provide the modifications to the first term.
After doing some tedious calculation we have found the following
results for $\mathcal{A}_{CP}(s)$ and $\mathcal{A}^i_{CP}(s)$
\begin{eqnarray}
\mathcal{A}_{CP}(s)=\frac{-2\mathcal{I}m(\Lambda_{sb})\mathcal{Q}(s)}{\mathcal{M}_1+2\mathcal{I}m(\Lambda_{sb})\mathcal{Q}(s)},\label{unpolcp}
\\
\mathcal{A}^i_{CP}(s)=\frac{-2\mathcal{I}m(\Lambda_{sb})\mathcal{Q}^i(s)}{\mathcal{M}_1+2\mathcal{I}m(\Lambda_{sb})\mathcal{Q}^i(s)},\label{polcp}
\end{eqnarray}
with $i=L,N\text{ or }T$. The explicit expressions of
$\mathcal{Q}(s)$ and the $\mathcal{Q}^i(s)$ are given below
\begin{eqnarray}
\mathcal{Q}(s)&=&\mathcal{H}_1\mathcal{I}m(C_7C_{9}^{\ast \prime})+\mathcal{H}_2\mathcal{I}m(C_9^{\ast}C_{9}^{\prime}),\notag\\
\mathcal{Q}^L(s)&=&\mathcal{H}_3\{\mathcal{I}m(C_{10}^{\prime} C_{9}^{\ast})+\mathcal{I}m(C_{9}^{\ast \prime}C_{10})\},\notag\\
\mathcal{Q}^N(s)&=&\mathcal{H}_4\mathcal{I}m(C_7C_{9}^{\ast \prime})+\mathcal{H}_5\{\mathcal{I}m(C_{10}C_{9}^{\ast \prime})\},\notag\\
&&+\mathcal{H}_6\mathcal{I}m(C_{9}^{\ast}C_{9}^{\prime}),\notag\\
\mathcal{Q}^T(s)&=&\frac{\beta}{2}\mathcal{H}_6\{\mathcal{R}e(C_{10}C_{9}^{\ast \prime})+\mathcal{R}e(C_{10}^{\prime} C_{9}^{\ast})\},\label{Qfunc}
\end{eqnarray}
The functions $\mathcal{H}_1,\ldots,\mathcal{H}_6$ can be written as
\begin{eqnarray}
\mathcal{H}_1&=&\frac{64}{3M_{\phi}^2s}M_{B_s}\bigg[(\mathcal{D}_3\mathcal{D}_5+\mathcal{D}_6\mathcal{D}_4)\lambda(M_{\phi}^2-M_{B_s}^2+s)\notag\\
&&+(\mathcal{D}_3\mathcal{D}_4+\mathcal{D}_5\mathcal{D}_6)\lambda+(3\mathcal{D}_3\mathcal{D}_4+2\mathcal{D}_1F_1(s)\lambda)\bigg],\notag
\\
\mathcal{H}_2&=&\frac{32(2m_l+s)^2}{3M_{\phi}^2s}\bigg[2\mathcal{D}_3\mathcal{D}_6\lambda(M_{\phi}^2-M_{B_s}^2+s)\notag
\\
&&+8\mathcal{D}_1^2M_{\phi}^2s\lambda+\mathcal{D}_3^2(12M_{\phi}^2s+\lambda)+\mathcal{D}_6^2\lambda^2\bigg],\notag
\\
\mathcal{H}_3&=&\frac{32\beta}{3M_{\phi}^2}\bigg[2\lambda(M_{B_s}^2-M_{\phi}^2-s)\mathcal{D}_3\mathcal{D}_6\notag
\\
&&-\mathcal{D}_3^2(12M_{\phi}^2s+\lambda)-\lambda(8M_{\phi}^2s\mathcal{D}_1^2+\lambda\mathcal{D}_6^2)\bigg],\label{Hfunctions}
\\
\mathcal{H}_4&=&\frac{128\pi}{\sqrt{s}}m_l
M_{B_s}\sqrt{\lambda}\bigg[F_1(s)\mathcal{D}_3+\mathcal{D}_1\mathcal{D}_4\bigg],\notag
\\
\mathcal{H}_5&=&\frac{8\pi
m_l\sqrt{\lambda}}{M_{\phi}^2\sqrt{s}}\bigg(\mathcal{D}_6(M_{\phi}^2-M_{B_s}^2)+\mathcal{D}_3-2\sqrt{s}\mathcal{D}_7\bigg)\notag
\\
&&\times\bigg(\mathcal{D}_3(M_{\phi}^2-M_{B_s}^2+s)+\mathcal{D}_6\lambda\bigg),\notag
\\
\mathcal{H}_6&=&128m_l\pi\sqrt{\lambda
s}\mathcal{D}_1\mathcal{D}_3.\notag
\end{eqnarray}
Here we would like to mention that the functions $\mathcal{H}_{4},\mathcal{H}_{5}$ and $\mathcal{H}_{6}$ are proportional to the
mass of final state lepton, therefore, their contribution is small when we have $\mu$'s as final state leptons compared to the one
when we have $\tau$'s.

\section{Numerical Analysis}
In this section we will give the phenomenological analysis of the polarized branching ratio, when final state $\phi$ meson is
longitudinal (transverse) polarized $(\mathcal{BR}_L(\mathcal{BR}_T))$, as well as of the unpolarized and polarized $CP$
violation asymmetries for $B_{s}\to \phi \ell^{+} \ell^{-}$ decay. In order to see the impact of new $Z^{\prime}$ boson on the these physical observables, first we have summarized the numerical values of various input parameters
such as masses of particles, life times, CKM matrix
elements etc., in Table II, while the values of Wilson
coefficients in the SM are displayed in Table III. The most important input parameters are the form factors
which are the non perturbative quantities and for them we rely on the Light
Cone Sum Rule (LCSR) approach. The numerical values of the LCSR form factors along with
the different fitting parameters \cite{PBALL} are summarized in Table I.
\begin{table}[ht]
\caption{Default values of input parameters used in the calculations.}
\label{input}\centering
\begin{tabular}{c}
\hline\hline
$M_{B_s}=5.366$ GeV, $m_{b}=4.28$ GeV, $m_{s}=0.13$ GeV, \\
$m_{\mu}=0.105$ GeV, $m_{\tau}=1.77$ GeV, $f_{B_{s}}=0.25$ GeV, \\
$|V_{tb}V_{ts}^{\ast}|=45\times 10^{-3}$, $\alpha^{-1}=137$, $%
G_{F}=1.17\times 10^{-5}$ GeV$^{-2}$, \\
$\tau_{B}=1.54\times 10^{-12}$ sec, $M_{\phi}=1.020$ GeV. \\ \hline\hline
\end{tabular}%
\end{table}

\begin{table*}[ht]
\centering \caption{The Wilson coefficients $C_{i}^{\mu}$ at the
scale $\mu\sim m_{b}$ in the SM \cite{LCSR}.}
\begin{tabular}{cccccccccc}
\hline\hline
$C_{1}$&$C_{2}$&$C_{3}$&$C_{4}$&$C_{5}$&$C_{6}$&$C_{7}$&$C_{9}$&$C_{10}$
\\ \hline
 \ \  1.107 \ \  & \ \  -0.248 \ \  & \ \  -0.011 \ \  & \ \  -0.026 \ \  & \ \  -0.007 \ \  & \ \  -0.031 \ \  & \ \  -0.313 \ \  & \ \  4.344 \ \  & \ \  -4.669 \ \  \\
\hline\hline
\end{tabular}
\label{wc table}
\end{table*}

Now the next step is to collect the values of the $Z^\prime$ couplings and in this regard,
there are some severe constraints from different
inclusive and exclusive $B$ decays \cite{ConstrainedZPC1}.
These numerical values of coupling parameters of $Z^\prime$ model
are recollected in Table 4, where $\mathcal{S}1$ and $\mathcal{S}2$ correspond to two
different fitting values for $B_s-\bar{B_{s}}$ mixing data by the
UTfit collaboration \cite{UTfit}.

Motivated from the latest results on the $CP$-violating phase $\phi^{L}_{S}$ and the like-sign dimuon
charge asymmetry $A^{b}_{SL}$ of the semileptonic decays given in \cite{cpv1,cpv2,cpv3,cpv4}, a detailed study
has been performed by Li et al. \cite{cpv5}. The main emphasis of the study is to check if a simultaneous explanation
for all mixing observables, especially of like-sign dimuon asymmetry  $A^{b}_{SL}$, could be made in $Z^{\prime}$ model.
It has been found that it is not possible to accommodate all the data simultaneously and the new constraints on the $CP$-violating phase $\phi_{S}$ and $|\mathcal{B}_{sb}|$ are obtained from $\Delta M_{S}$, $\phi_{S}$, $\Delta\Gamma_{S}$ data. In addition the constraints on $S^{L}_{\ell \ell}$ and $S^{R}_{\ell \ell}$ are obtained from the analysis of $B\to X_{s}\mu^+\mu^-$ \cite{newcon1a}, $B\to K^{*}\mu^+\mu^-$ \cite{newcon1,newcon2} and $B\to \mu^+\mu^-$ \cite{newcon3}. In the forthcoming study it is named as scenario $\mathcal{S}3$. The corresponding numerical values are chosen from \cite{cpv5, newcon4} and these are summarized in the Table IV.
\begin{table*}[ht]
\centering \caption{The numerical values of the $Z^\prime$
parameters \cite{ConstrainedZPC1,UTfit,cpv5,newcon4}.}
\begin{tabular}{cccccc}
\hline\hline \ \ &  $|\mathcal{B}_{sb}|\times10^{-3}$ \ \  & \ \ $
\phi_{sb}(in Degree)$ \ \  & \ \ $S_{LL}\times10^{-2}$ \ \  & \ \
$D_{LL}\times10^{-2}$
\\ \hline
 \ \  $\mathcal{S}1$ \ \  & \ \  $1.09\pm0.22$ \ \  & \ \  $-72\pm7$ \ \  & \ \  $-2.8\pm3.9$ \ \  & \ \  $-6.7\pm2.6$ \\
\ \  $\mathcal{S}2$ \ \  & \ \  $2.20\pm0.15$ \ \  & \ \  $-82\pm4$ \ \  & \ \  $-1.2\pm1.4$ \ \  & \ \  $-2.5\pm0.9$  \\
\ \  $\mathcal{S}3$ \ \  & \ \  $4.0\pm1.5$ \ \  & \ \  $150\pm10$ or $(-150\pm10)$ \ \  & \ \  $0.8$ \ \  & \ \  $-2.6$  \\
\hline\hline
\end{tabular}
\label{ZP table}
\end{table*}

\begin{figure}[tbp]
\begin{tabular}{cc}
\epsfig{file=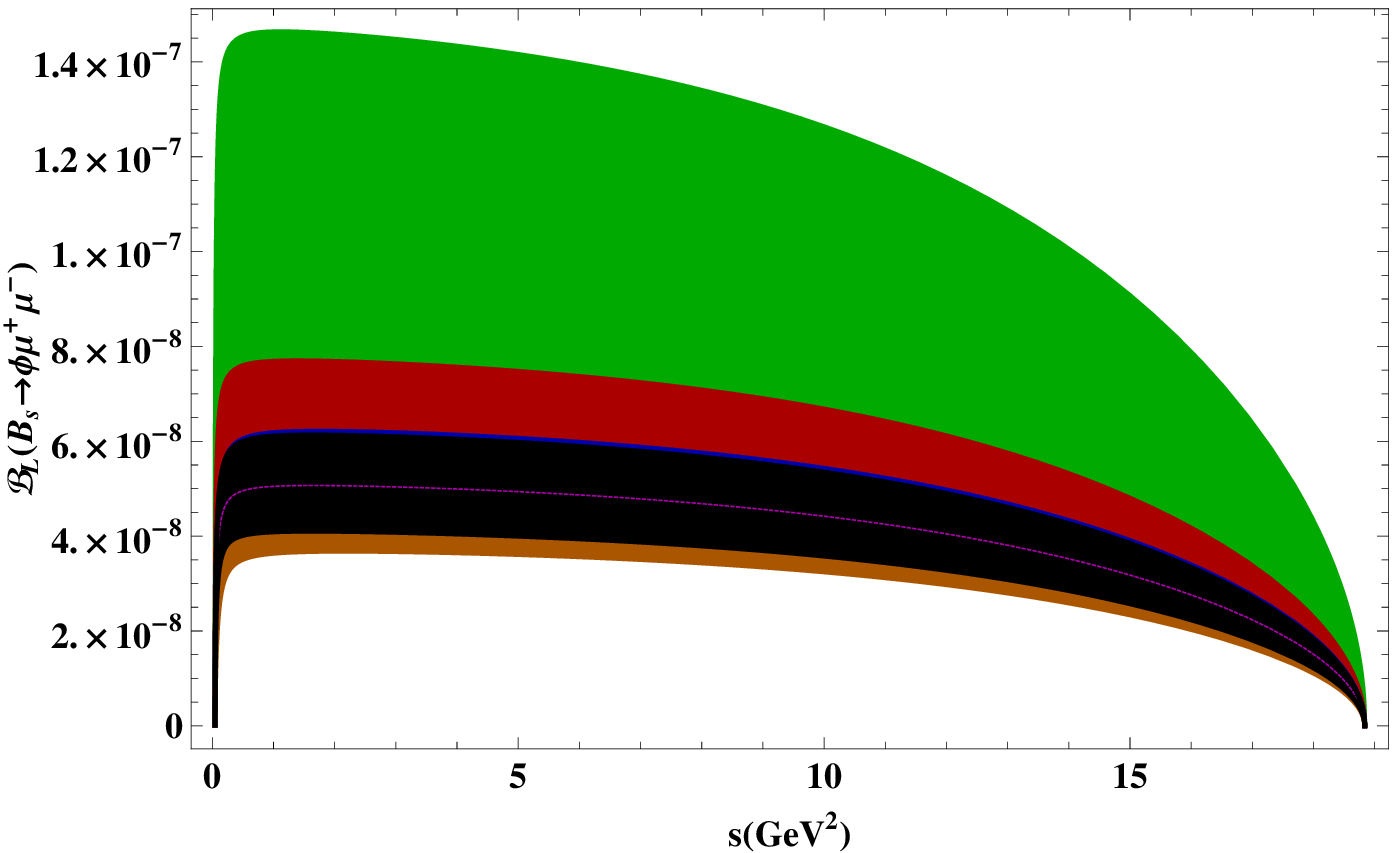,width=0.51\linewidth,clip=a} \put (-100,150){(a)} & %
\epsfig{file=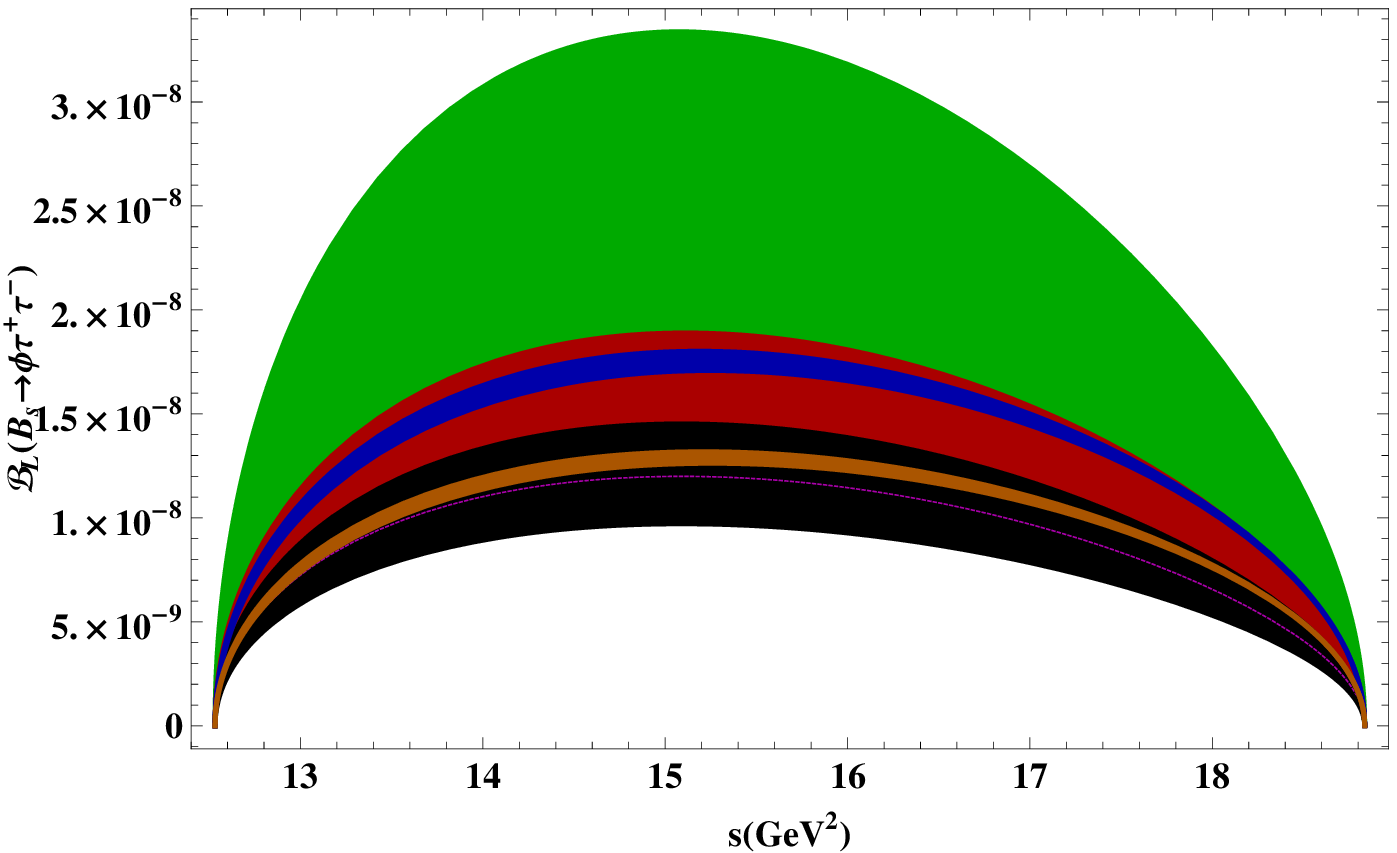,width=0.51\linewidth,clip=b} \put (-100,150){(b)}%
\end{tabular}%
\caption{Longitudinal polarized branching ratio $\mathcal{BR}_L$ as a function of the square of momentum $s$ for $B_s\to \phi \mu^+ \mu^- (\tau^+\tau^-)$ for scenarios $\mathcal{S}1$,
$\mathcal{S}2$ and $\mathcal{S}3$. The Green and Red colors correspond to the $\mathcal{S}1$ and $\mathcal{S}2$, respectively. The blue and orange colors show the $\mathcal{S}3$. The band in each case depicts the variations of $\phi_{sb}$ in respective
scenarios. The black color corresponds to the SM results where the band is due to uncertainties in different input parameters.} \label{LBR}
\end{figure}

\begin{figure}[tbp]
\begin{tabular}{cc}
\epsfig{file=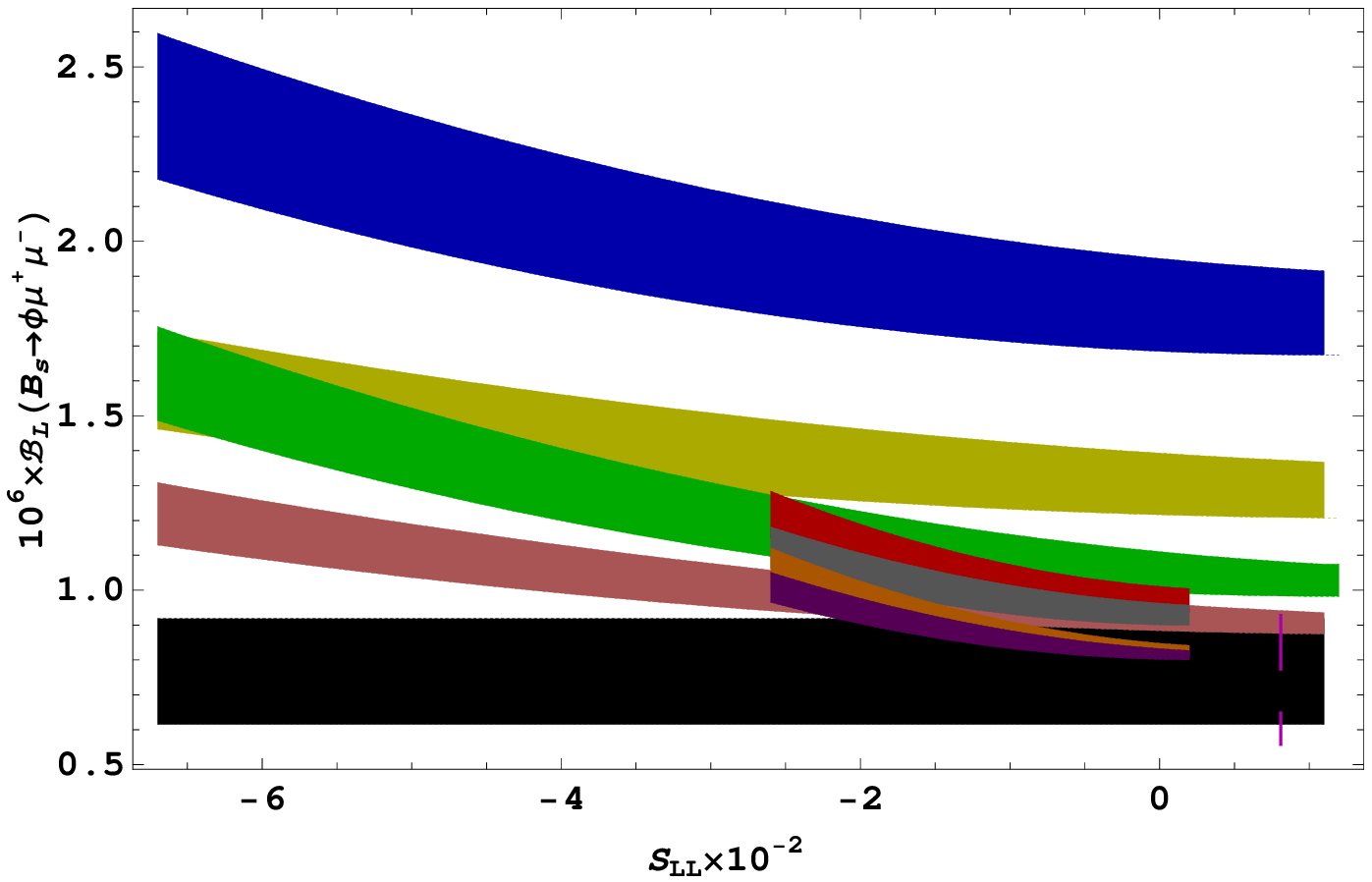,width=0.51\linewidth,clip=a} \put (-100,150){(a)} & %
\epsfig{file=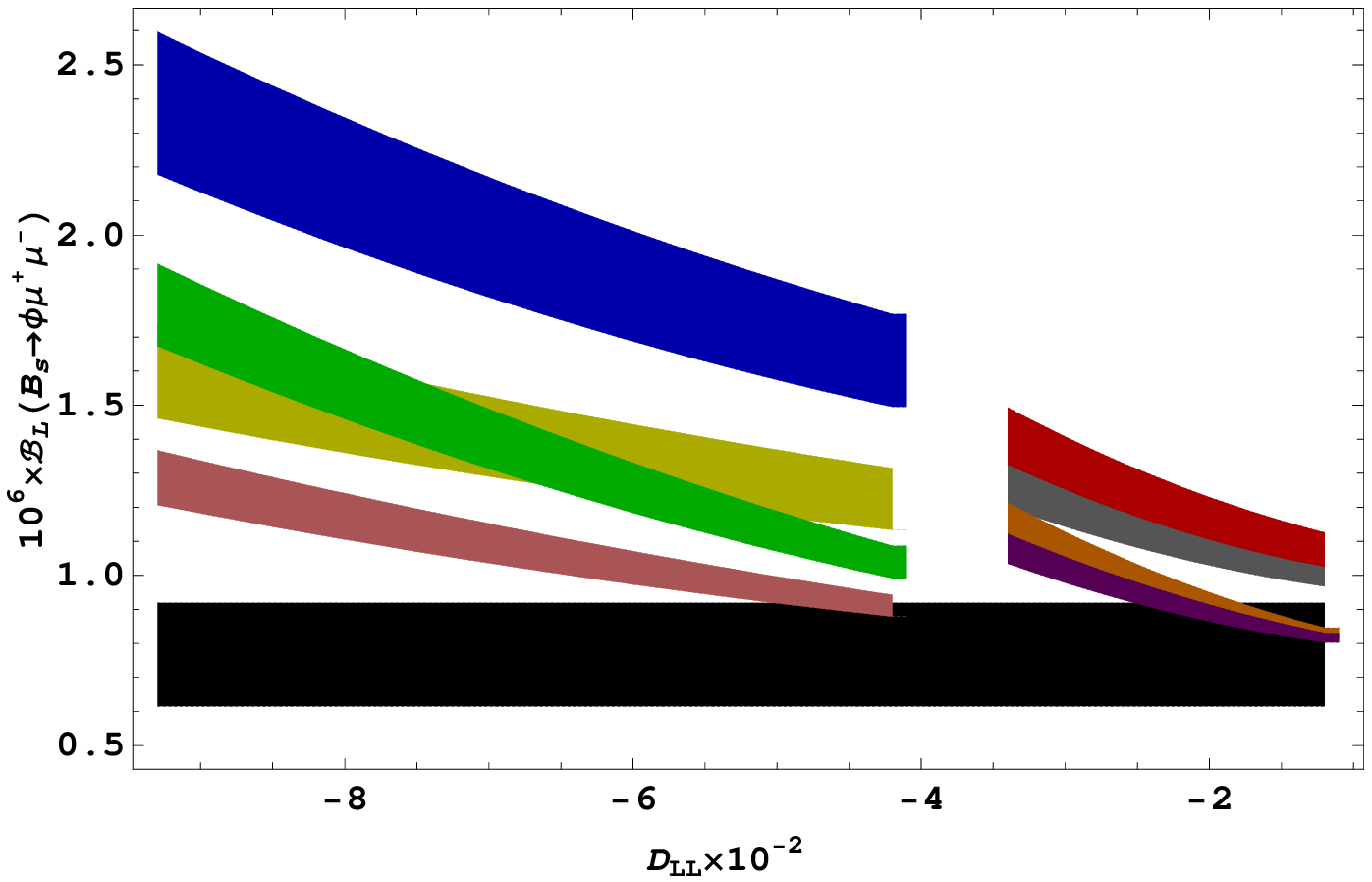,width=0.51\linewidth,clip=b} \put (-100,150){(b)} \\
\epsfig{file=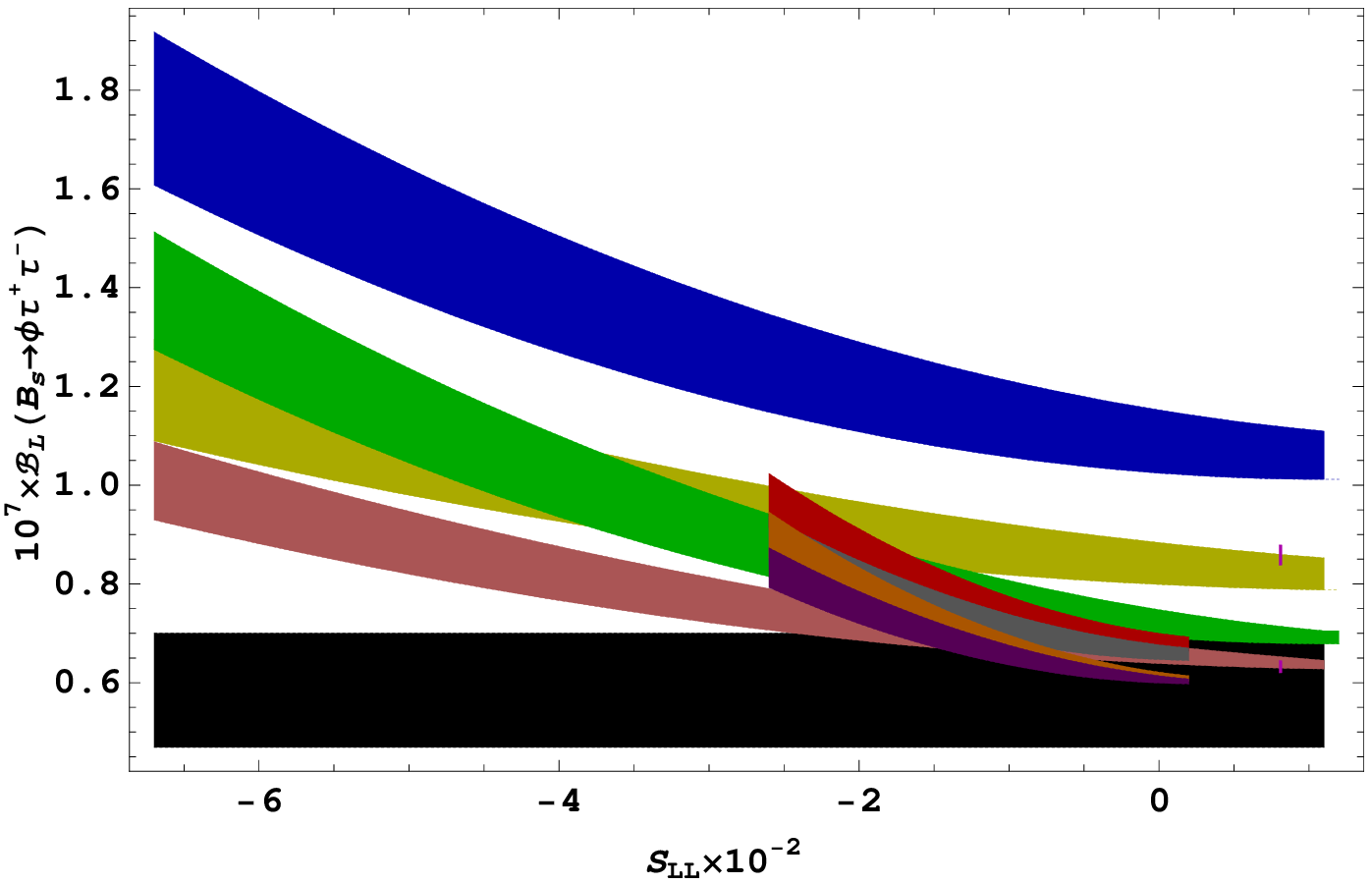,width=0.51\linewidth,clip=c} \put (-100,150){(c)} & %
\epsfig{file=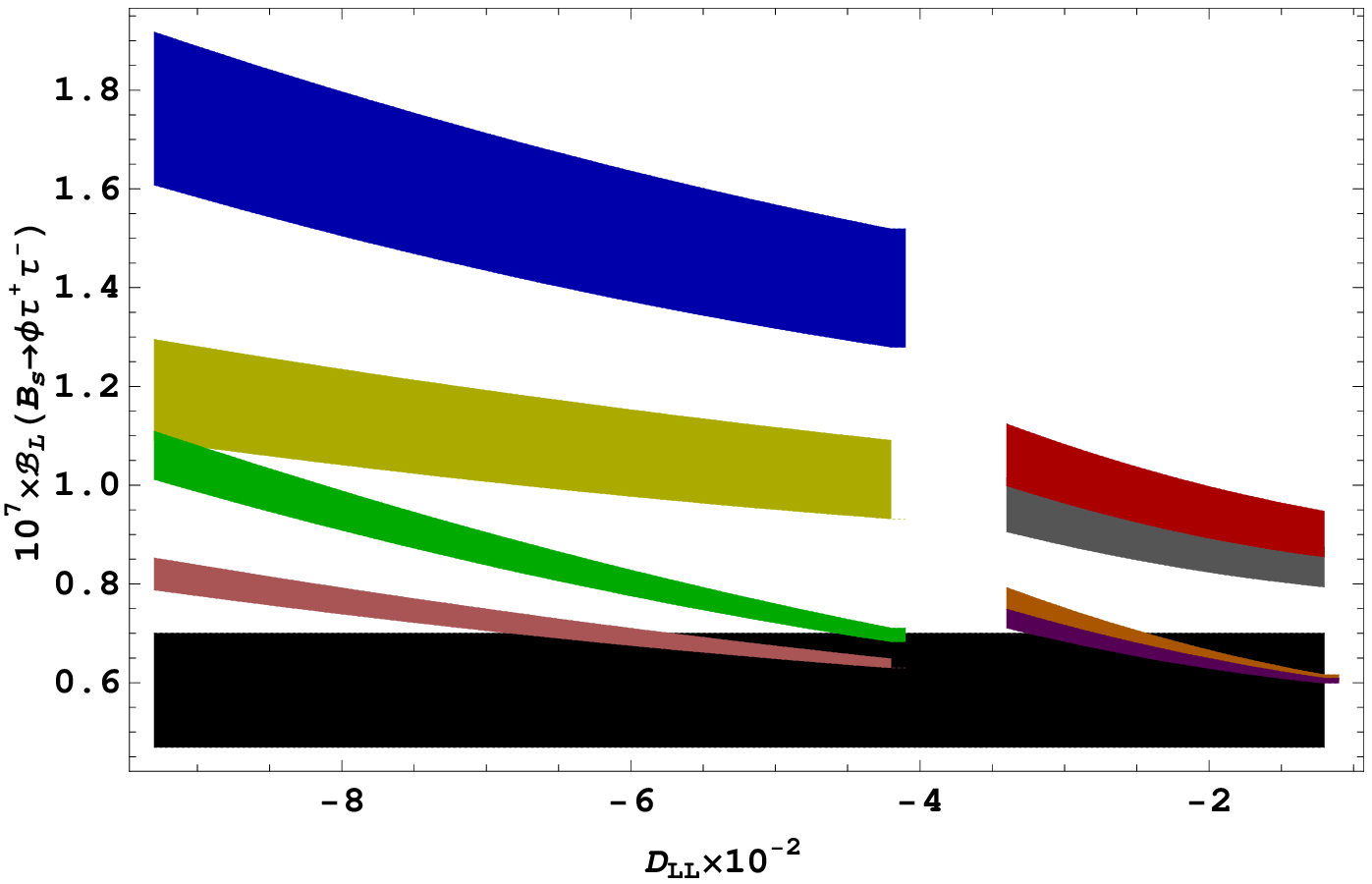,width=0.51\linewidth,clip=d} \put (-100,150){(d)}%
\end{tabular}%
\caption{Longitudinal polarized branching ratio $\mathcal{BR}_L$ as a function of the square of momentum $D_{LL}$ and $S_{LL}$ for $B_s\to \phi \mu^+ \mu^- (\tau^+\tau^-)$ for scenarios $\mathcal{S}1$ and
$\mathcal{S}2$. The Blue, Yellow, Green and Dark Pink colors correspond to the $\mathcal{S}1$ where as the
other colors are for the $\mathcal{S}2$. The vertical magenta color bars corresponding to the $\mathcal{S}3$ scenario.
The band in each case depicts the variations of $\phi_{sb}$ in respective
scenario. The black band corresponds to the SM results where the band is due to uncertainty in different input parameters.}
\label{BRLMT}
\end{figure}

\begin{figure}[tbp]
\begin{tabular}{cc}
\epsfig{file=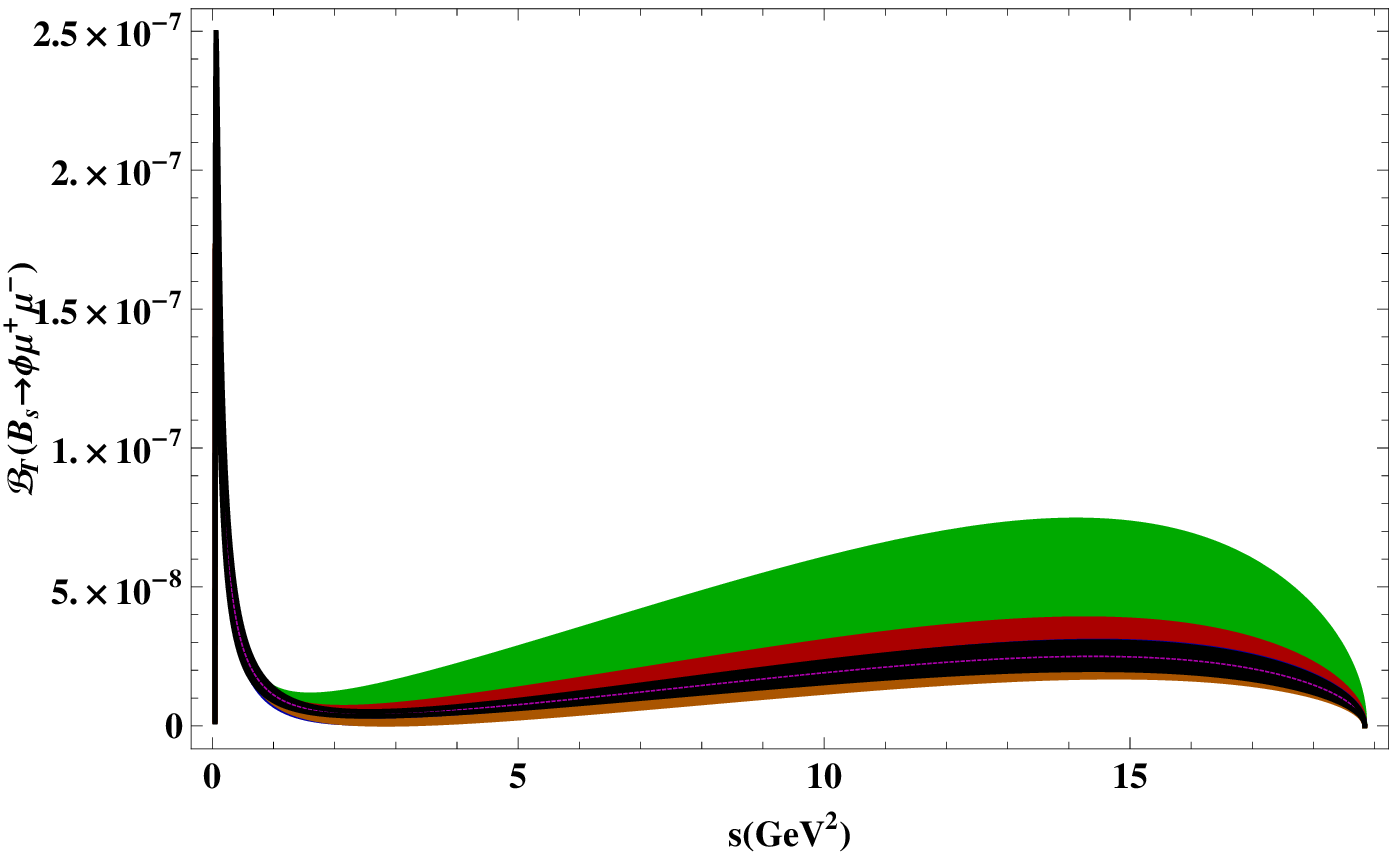,width=0.51\linewidth,clip=a} \put (-100,150){(a)} & %
\epsfig{file=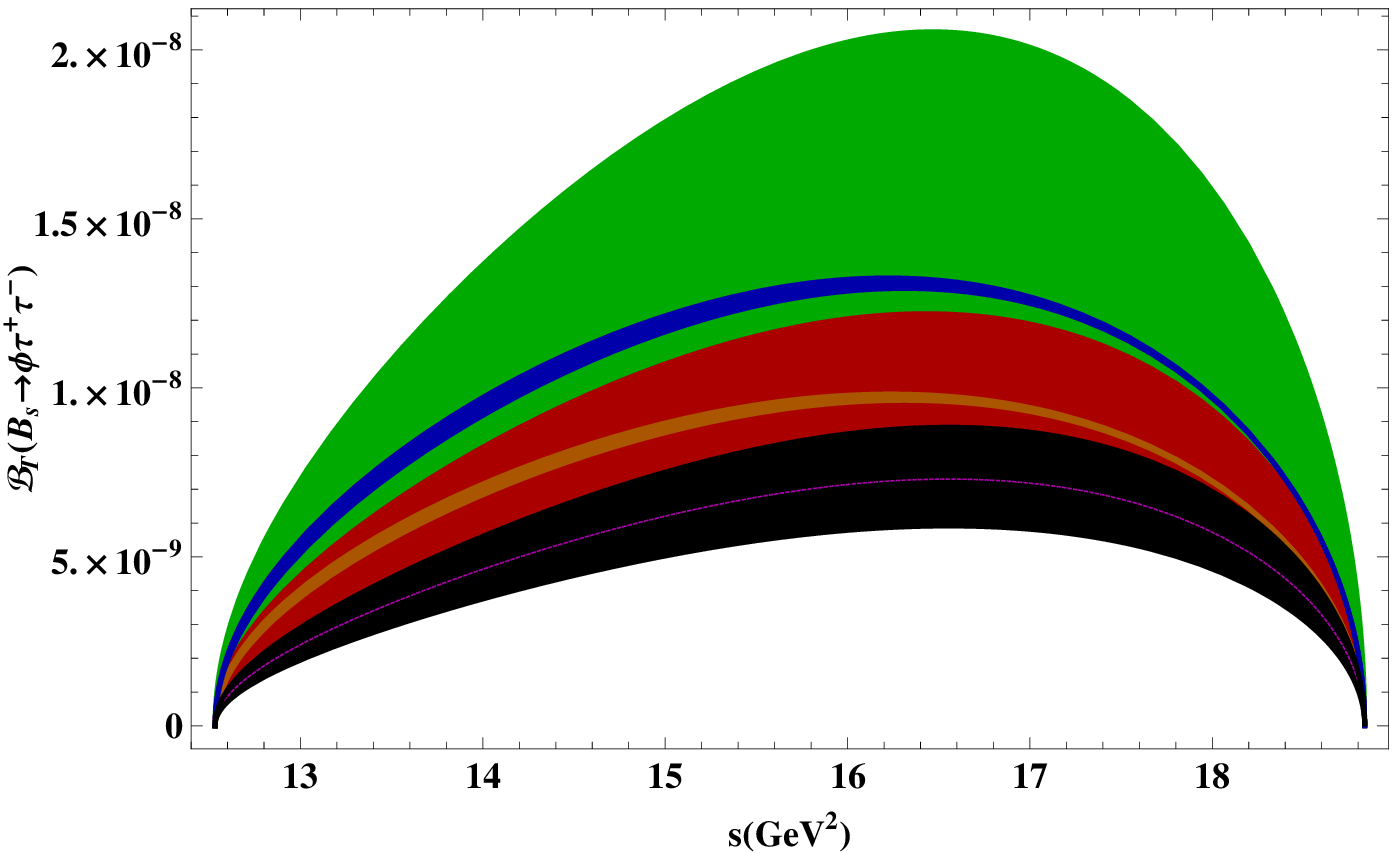,width=0.51\linewidth,clip=b} \put (-100,150){(b)}%
\end{tabular}%
\caption{Transverse polarized branching ratio $\mathcal{BR}_T$ s a function of the square of momentum $s$ for $B_s\to \phi \mu^+ \mu^- (\tau^+\tau^-)$ for scenarios $\mathcal{S}1$,
$\mathcal{S}2$ and $\mathcal{S}3$. The Green and Red colors correspond to the $\mathcal{S}1$ and $\mathcal{S}2$, respectively. The blue and orange colors show the $\mathcal{S}3$. The band in each case depicts the variations of $\phi_{sb}$ in respective
scenarios. The black color corresponds to the SM results where the band is due to uncertainties in different input parameters.} \label{TBR}
\end{figure}

\begin{figure}[tbp]
\begin{tabular}{cc}
\epsfig{file=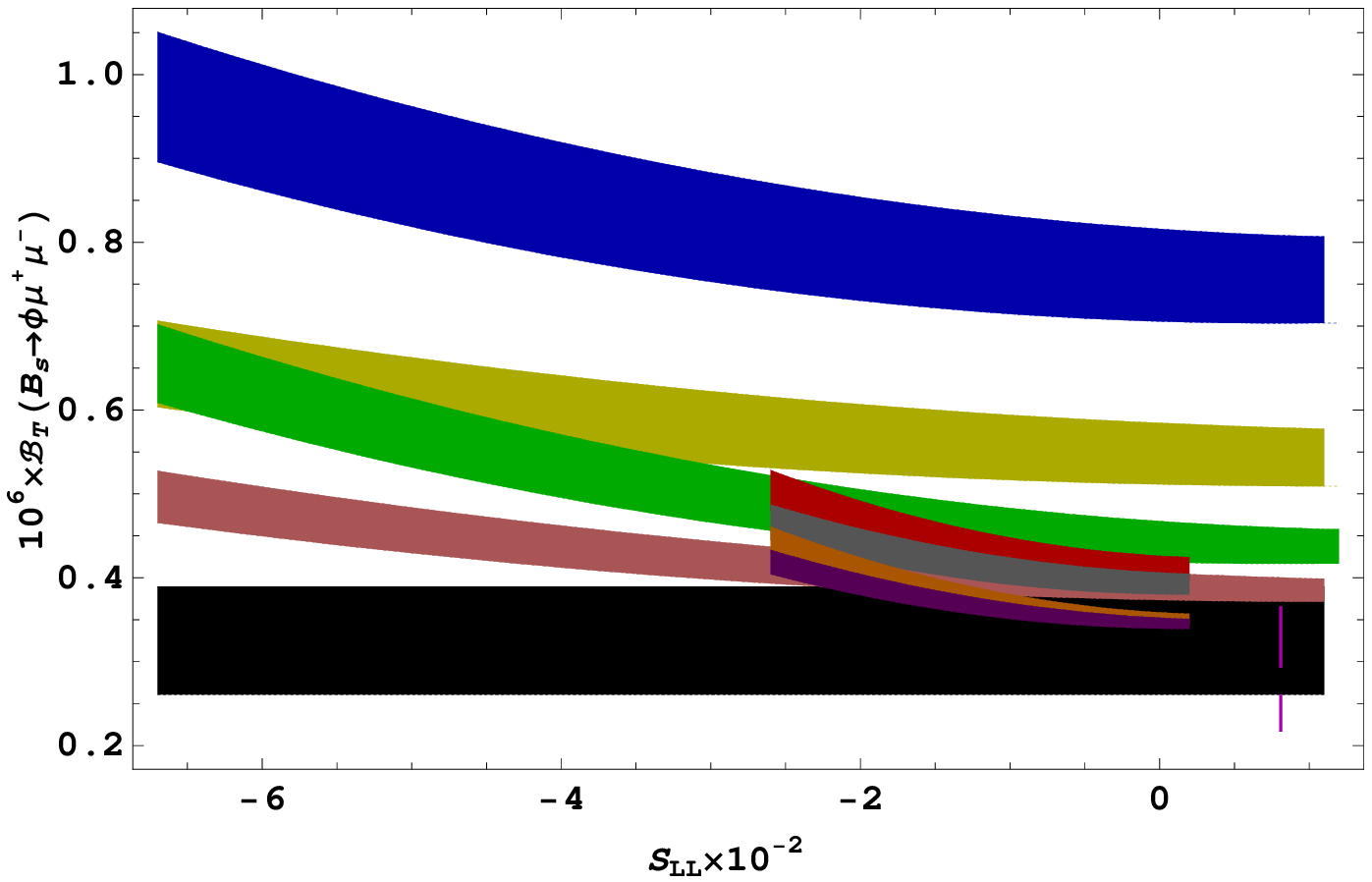,width=0.51\linewidth,clip=a} \put (-100,150){(a)} & %
\epsfig{file=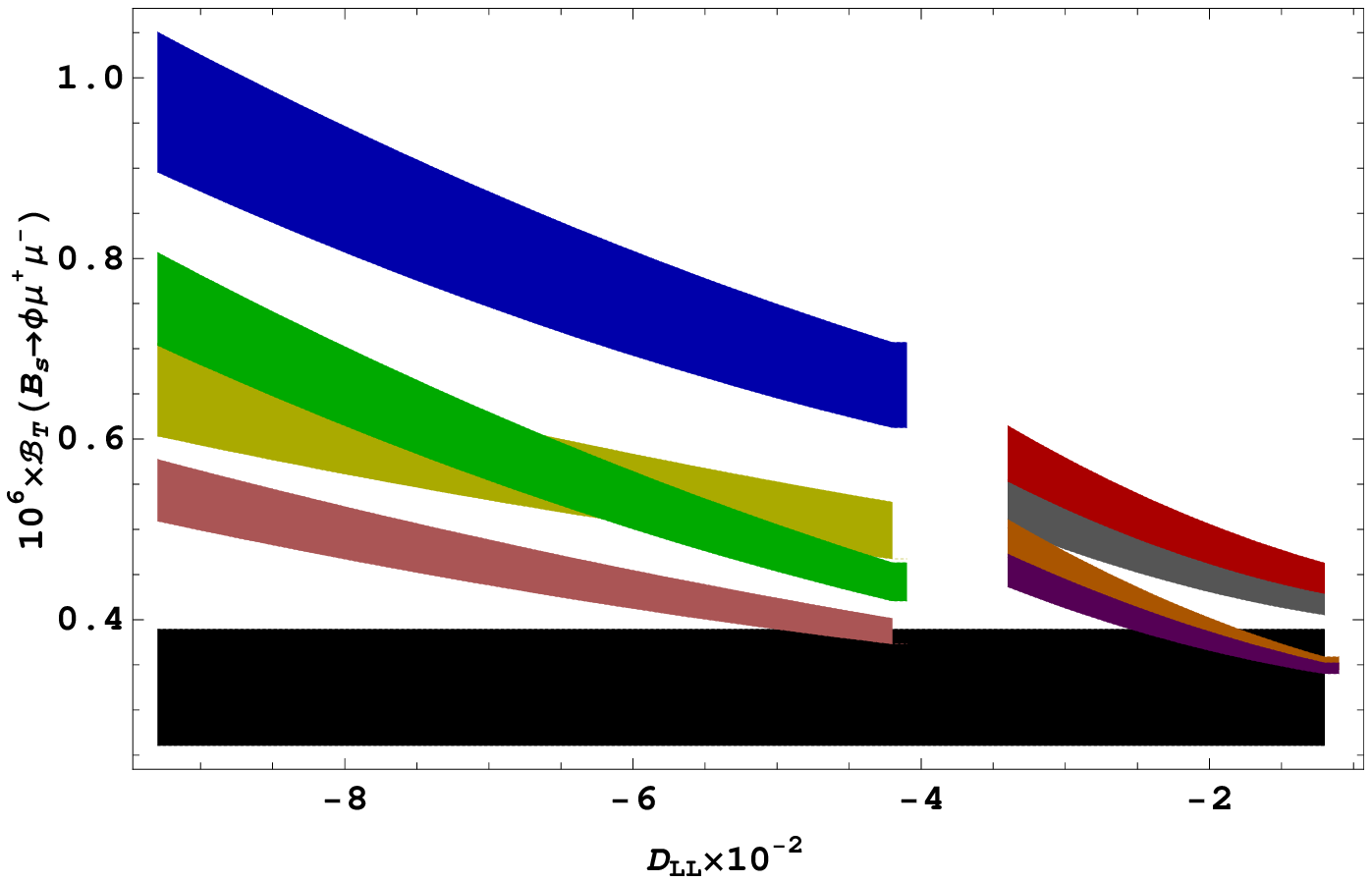,width=0.51\linewidth,clip=b} \put (-100,150){(b)} \\
\epsfig{file=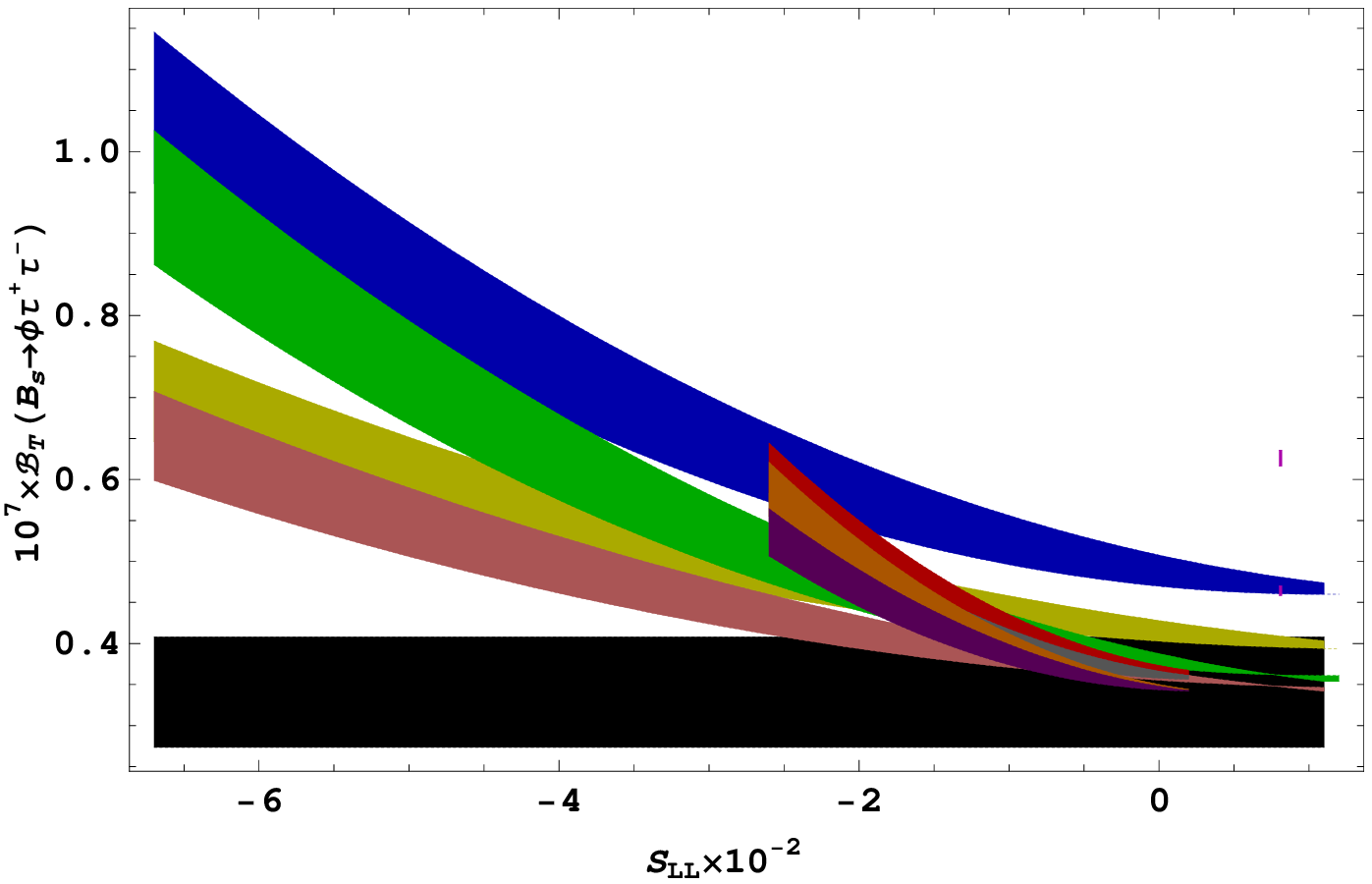,width=0.51\linewidth,clip=c} \put (-100,150){(c)} & %
\epsfig{file=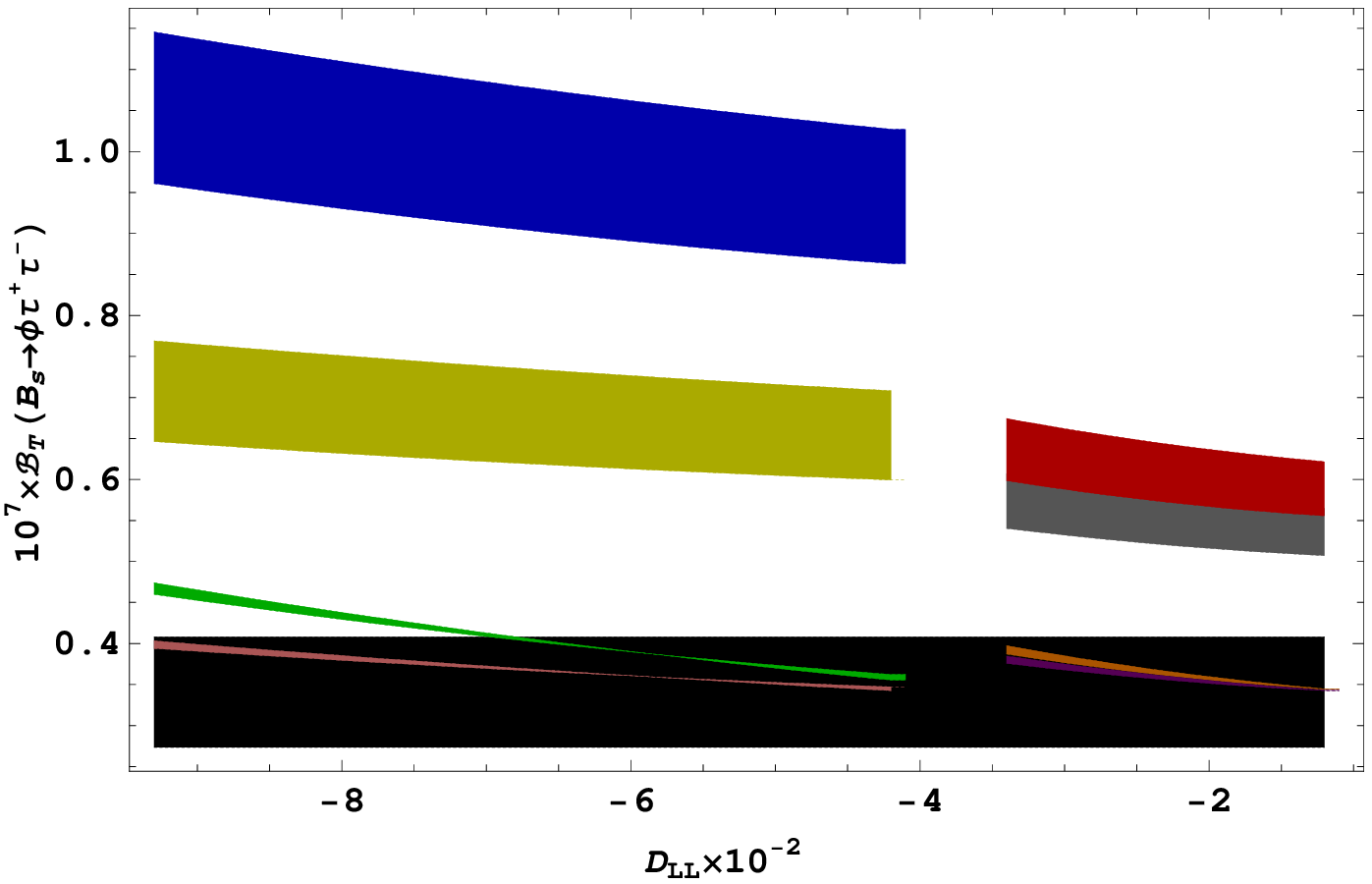,width=0.51\linewidth,clip=d} \put (-100,150){(d)}%
\end{tabular}%
\caption{The transverse lepton polarization asymmetry for the
$B_{s} \to \phi l^+l^-$ ($l=\protect\mu, \protect\tau$) decays as functions of
$Z^{\prime}$ parameters. The legends are same as in Fig.2. }\label{BRTZP}
\end{figure}

Just to mention again, $\mathcal{B}_{sb}=|\mathcal{B}_{sb}|e^{-i\phi_{sb}}$ is the off diagonal left handed coupling of
$Z^\prime$ boson with quarks and $\phi_{sb}$ corresponds to a new weak phase, whereas
$S_{LL}$ and $D_{LL}$ represent the combination of left and
right handed couplings of $Z^\prime$ with the leptons [c.f. Eq. (\ref{C9zdefinition})].
In order to fully scan the three scenarios, let us make a remark that with $D_{LL}\neq0$
depict the situation when the new physics comes only from the
modification in the Wilson coefficient $C_{10}$, while, the opposite
case, $S_{LL}\neq0$ , indicates that the new physics
is due to the change in
the Wilson coefficient $C_{9}$ [see Eq. (\ref{C9zdefinition})]. In Figs. \ref{LBR}$-$\ref{BRTZP} we have
displayed the results of the branching ratio when the final state meson $(\phi)$ is polarized. Figs. \ref{LBR} and \ref{TBR} represent
the cases where the $\mathcal{BR}_{L}$ and $\mathcal{BR}_T$ are plotted as a function of $s$ by taking the values of different
$Z^{\prime}$ parameters given in Table \ref{ZP table}. In Figs. \ref{BRLMT} and \ref{BRTZP} the average normalized polarized branching ratios, after integration on $s$,
as a function of $S_{LL}$ and $D_{LL}$ are depicted. In the same way, the averaged $CP$ violation asymmetries as a function of
$S_{LL}$ and $D_{LL}$ is shown in Figs. 5 $-$ 12.
The different color combinations along with the corresponding values of $Z^{\prime}$ parameters are summarize in the
Table V.
Likewise, in scenario $\mathcal{S}3$ the values of $Z^{\prime}$ parameters are summarized in Table IV
and their color codes in different figures are given in Eq. (\ref{Newcons}).
\begin{equation}
|\mathcal{B}_{sb}|=3\times 10^{-3}:\left\{
\begin{array}{c}
\phi _{sb}=160^{\circ }\text{, Red Dot} \\
\phi _{sb}=140^{\circ }\text{, Blue Dot} \\
\phi _{sb}=-140^{\circ }\text{, Green Dot} \\
\phi _{sb}=-160^{\circ }\text{, Gray Dot}%
\end{array}%
\right. \text{; }|\mathcal{B}_{sb}|=5\times 10^{-3}:\left\{
\begin{array}{c}
\phi _{sb}=160^{\circ }\text{, Orange Dot} \\
\phi _{sb}=140^{\circ }\text{, Yellow Dot} \\
\phi _{sb}=-140^{\circ }\text{, Pink Dot} \\
\phi _{sb}=-160^{\circ }\text{, Purple Dot}%
\end{array}%
\right.   \label{Newcons}
\end{equation}

\begin{table*}[ht] \centering \caption{Color bands for the Figs. 1 $-$ 12 $\langle\mathcal{BR}_{L,T}\rangle$, $\langle\mathcal{A_{CP}}\rangle$ and
$\langle\mathcal{A}^{i}_{CP}\rangle$ vs $S_{LL}$ and $D_{LL}$ for scenarios $\mathcal{S}1$ and $\mathcal{S}2$.}
\begin{tabular}{|c|c|c|c|c|} \cline{1-5}
\multicolumn{1}{|c|}{\multirow{2}{*}{Color Region}} &
\multicolumn{1}{|c|}{\multirow{2}{*}{$\phi_{sb}$}} &
\multicolumn{1}{|c|}{\multirow{2}{*}{$|\mathcal{B}_{sb}|\times 10^{-3}$}} &
\multicolumn{1}{|c|}{\multirow{1}{*}{$\langle\mathcal{BR}_{L,T}\rangle$, $\langle\mathcal{A_{CP}}\rangle$ and
$\langle\mathcal{A}^{i}_{CP}\rangle$ vs $S_{LL}$}} &
\multicolumn{1}{|c|}{\multirow{1}{*}{$\langle\mathcal{BR}_{L,T}\rangle$, $\langle\mathcal{A_{CP}}\rangle$ and
$\langle\mathcal{A}^{i}_{CP}\rangle$ vs $D_{LL}$}} \\
\multicolumn{1}{|c|}{\multirow{1}{*}{}} &
\multicolumn{1}{|c|}{\multirow{1}{*}{}} &
\multicolumn{1}{|c|}{\multirow{1}{*}{}} &
\multicolumn{1}{|c|}{\multirow{1}{*}{$D_{LL}\times 10^{-2}$}} &
\multicolumn{1}{|c|}{\multirow{1}{*}{$S_{LL}\times 10^{-2}$}} \\
\cline{1-5}\multicolumn{1}{|c|}{\multirow{2}{*}{Blue}} &
\multicolumn{1}{|c|}{\multirow{1}{*}{$-79^\circ$}} &
\multicolumn{1}{|c|}{\multirow{2}{*}{+1.31}} &
\multicolumn{1}{|c|}{\multirow{2}{*}{-9.3}} &
\multicolumn{1}{|c|}{\multirow{2}{*}{-6.7}} \\
\multicolumn{1}{|c|}{\multirow{1}{*}{}} &
\multicolumn{1}{|c|}{\multirow{1}{*}{$-65^\circ$}} &
\multicolumn{1}{|c|}{\multirow{1}{*}{}} &
\multicolumn{1}{|c|}{\multirow{1}{*}{}} &
\multicolumn{1}{|c|}{\multirow{1}{*}{}} \\
\cline{1-5} \multicolumn{1}{|c|}{\multirow{2}{*}{Red}} &
\multicolumn{1}{|c|}{\multirow{1}{*}{$-86^\circ$}} &
\multicolumn{1}{|c|}{\multirow{2}{*}{+2.35}} &
\multicolumn{1}{|c|}{\multirow{2}{*}{-2.34}} &
\multicolumn{1}{|c|}{\multirow{2}{*}{-2.6}} \\
\multicolumn{1}{|c|}{\multirow{1}{*}{}} &
\multicolumn{1}{|c|}{\multirow{1}{*}{$-78^\circ$}} &
\multicolumn{1}{|c|}{\multirow{1}{*}{}} &
\multicolumn{1}{|c|}{\multirow{1}{*}{}} &
\multicolumn{1}{|c|}{\multirow{1}{*}{}} \\
\cline{1-5} \multicolumn{1}{|c|}{\multirow{2}{*}{Yellow}} &
\multicolumn{1}{|c|}{\multirow{1}{*}{$-79^\circ$}} &
\multicolumn{1}{|c|}{\multirow{2}{*}{+0.87}} &
\multicolumn{1}{|c|}{\multirow{2}{*}{-9.3}} &
\multicolumn{1}{|c|}{\multirow{2}{*}{-6.7}} \\
\multicolumn{1}{|c|}{\multirow{1}{*}{}} &
\multicolumn{1}{|c|}{\multirow{1}{*}{$-65^\circ$}} &
\multicolumn{1}{|c|}{\multirow{1}{*}{}} &
\multicolumn{1}{|c|}{\multirow{1}{*}{}} &
\multicolumn{1}{|c|}{\multirow{1}{*}{}} \\
\cline{1-5} \multicolumn{1}{|c|}{\multirow{2}{*}{Black}} &
\multicolumn{1}{|c|}{\multirow{1}{*}{$-86^\circ$}} &
\multicolumn{1}{|c|}{\multirow{2}{*}{+2.05}} &
\multicolumn{1}{|c|}{\multirow{2}{*}{-2.34}} &
\multicolumn{1}{|c|}{\multirow{2}{*}{-2.6}} \\
\multicolumn{1}{|c|}{\multirow{1}{*}{}} &
\multicolumn{1}{|c|}{\multirow{1}{*}{$-78^\circ$}} &
\multicolumn{1}{|c|}{\multirow{1}{*}{}} &
\multicolumn{1}{|c|}{\multirow{1}{*}{}} &
\multicolumn{1}{|c|}{\multirow{1}{*}{}} \\
\cline{1-5} \multicolumn{1}{|c|}{\multirow{2}{*}{Green}} &
\multicolumn{1}{|c|}{\multirow{1}{*}{$-79^\circ$}} &
\multicolumn{1}{|c|}{\multirow{2}{*}{+1.31}} &
\multicolumn{1}{|c|}{\multirow{2}{*}{-4.1}} &
\multicolumn{1}{|c|}{\multirow{2}{*}{+1.1}} \\
\multicolumn{1}{|c|}{\multirow{1}{*}{}} &
\multicolumn{1}{|c|}{\multirow{1}{*}{$-65^\circ$}} &
\multicolumn{1}{|c|}{\multirow{1}{*}{}} &
\multicolumn{1}{|c|}{\multirow{1}{*}{}} &
\multicolumn{1}{|c|}{\multirow{1}{*}{}} \\
\cline{1-5} \multicolumn{1}{|c|}{\multirow{2}{*}{Brown}} &
\multicolumn{1}{|c|}{\multirow{1}{*}{$-86^\circ$}} &
\multicolumn{1}{|c|}{\multirow{2}{*}{+2.35}} &
\multicolumn{1}{|c|}{\multirow{2}{*}{-1.16}} &
\multicolumn{1}{|c|}{\multirow{2}{*}{+0.2}} \\
\multicolumn{1}{|c|}{\multirow{1}{*}{}} &
\multicolumn{1}{|c|}{\multirow{1}{*}{$-78^\circ$}} &
\multicolumn{1}{|c|}{\multirow{1}{*}{}} &
\multicolumn{1}{|c|}{\multirow{1}{*}{}} &
\multicolumn{1}{|c|}{\multirow{1}{*}{}} \\
\cline{1-5} \multicolumn{1}{|c|}{\multirow{2}{*}{Pink}} &
\multicolumn{1}{|c|}{\multirow{1}{*}{$-79^\circ$}} &
\multicolumn{1}{|c|}{\multirow{2}{*}{+0.87}} &
\multicolumn{1}{|c|}{\multirow{2}{*}{-4.1}} &
\multicolumn{1}{|c|}{\multirow{2}{*}{+1.1}} \\
\multicolumn{1}{|c|}{\multirow{1}{*}{}} &
\multicolumn{1}{|c|}{\multirow{1}{*}{$-65^\circ$}} &
\multicolumn{1}{|c|}{\multirow{1}{*}{}} &
\multicolumn{1}{|c|}{\multirow{1}{*}{}} &
\multicolumn{1}{|c|}{\multirow{1}{*}{}} \\
\cline{1-5} \multicolumn{1}{|c|}{\multirow{2}{*}{Purple}} &
\multicolumn{1}{|c|}{\multirow{1}{*}{$-86^\circ$}} &
\multicolumn{1}{|c|}{\multirow{2}{*}{+2.05}} &
\multicolumn{1}{|c|}{\multirow{2}{*}{-1.16}} &
\multicolumn{1}{|c|}{\multirow{2}{*}{+0.2}} \\
\multicolumn{1}{|c|}{\multirow{1}{*}{}} &
\multicolumn{1}{|c|}{\multirow{1}{*}{$-78^\circ$}} &
\multicolumn{1}{|c|}{\multirow{1}{*}{}} &
\multicolumn{1}{|c|}{\multirow{1}{*}{}} &
\multicolumn{1}{|c|}{\multirow{1}{*}{}} \\
\cline{1-5}
\end{tabular}
\end{table*}

\textbf{Longitudinal polarized branching ratio $\mathcal{BR}_L$}
\begin{itemize}
\item In Figs. 1(a) and 1(b) we have plotted the branching ratio when final state $\phi$ is longitudinally polarized, named as
longitudinal polarized branching ratio $(\mathcal{BR}_L)$, as a function of $s$ for $\mu$ and $\tau$ as final state leptons'
in $B_s\to \phi \ell^+ \ell^-$ decay. By looking at Eq. (\ref{65j}) it can be seen that $\mathcal{BR}_L$ it is directly proportional
to the contributions coming from $Z^{\prime}$ in $C^{\prime}_{10}$ encoded in $\mathcal{K}_5,\mathcal{K}_6$ and $\mathcal{K}_7$. Apart from this
it also contains the terms that involve $C^{\prime}_9$ which comes in $\mathcal{K}_2$ and $\mathcal{K}_7$, where the later are $m_{l}$ suppressed.
In Fig. 1, we can see a significant enhancement in the $\mathcal{BR}_L$ for the maximum values of $Z^{\prime}$ parameters and the results are quite distinct from the SM both for $\mu$ and $\tau$ cases.

\item To see the explicit dependence on the $Z^{\prime}$ parameters we have integrated $\mathcal{BR}_L$ on $s$ and have
drawn it against the $D_{LL}$ and $S_{LL}$ in Figs. 2(a)-2(d). These graphs depict that for $\phi_{sb}=-79^{\circ},%
\mathcal{B}_{sb}=1.31\times 10^{-3}, D_{LL}=-9.3\times 10^{-2}$ and $S_{LL}=-6.7\times 10^{-2}$ in scenario $\mathcal{S}1$  (blue band), the increment in the $\mathcal{BR}_{L}$ is
around 3 times in case of $\mu$ and 2.5 times in case of $\tau$ leptons. By decreasing the values of
$D_{LL}$ and $S_{LL}$ the values of integrated $\mathcal{BR}_L$ decreases and Fig. 2 display this trend. Compared to the scenario $\mathcal{S}1$ the change in $\mathcal{BR}_L$ is small in $\mathcal{S}2$.

Keeping in view that in scenario $\mathcal{S}3$ the values of $S_{LL}$ and $D_{LL}$ are fixed, we plotted two vertical (magenta) bars which corresponds to the variation in the $\phi_{sb}$ and $\mathcal{B}_{sb}$. It can be seen from Figs. 2(a) and 2(c) that in certain range of parameters of $\mathcal{S}3$ the $Z^{\prime}$ boson effects are noticeable in $B_{s}\to \phi \tau^{+}\tau^{-}$ decay. Similar bars can be plotted in Figs. 2(b) and 2(d) but that do not add any new information, therefore, we will show $\mathcal{S}3$ contribution only when different asymmetries are plotted against $S_{LL}$.
\end{itemize}
\textbf{Transverse polarized branching ratio $\mathcal{BR}_T$}
\begin{itemize}
\item It can be noticed from Eq. (\ref{65k}) that transverse polarized branching ratio $(\mathcal{BR}_T)$ depends on the functions
$\mathcal{K}_1(s),\mathcal{K}_2(s),\mathcal{K}_4(s)$ and $\mathcal{K}_5(s)$ given in Eqs. (\ref{621a}), (\ref{621b}), (\ref{621d}) and (\ref{621e}), respectively. Here the first two functions $(\mathcal{K}_1(s),\mathcal{K}_2(s))$ depend on Wilson coefficients
$C^{eff}_7, C^{\prime}_9$ and the later two on $C^{\prime}_{10}$. Therefore, we are expecting quite visible hints of NP
coming from the the extra neutral boson $Z^{\prime}$ and Figs. 3(a) and 3(b), where $\mathcal{BR}_T$ is plotted as a function of
$s$, display this fact. Here, one can clearly distinguish between the values of $\mathcal{BR}_T$ calculated in SM and $Z^{\prime}$ scenarios,
$\mathcal{S}1$, $\mathcal{S}2$ and $\mathcal{S}3$.
\item To see how $\mathcal{BR}_T$ evolve with the parameters of $Z^\prime$ model, we have plotted the integrated $\mathcal{BR}_T$ as a function
of $S_{LL}$ and $D_{LL}$ in Figs. 4(a)-4(d). Just like $\mathcal{BR}_L$, $\mathcal{BR}_T$ becomes almost 3 times that of its SM value when
$\phi_{sb}=-79^{\circ}, %
\mathcal{B}_{sb}=1.31\times 10^{-3}, D_{LL}=-9.3\times 10^{-2}$ and $S_{LL}=-6.7\times 10^{-2}$ in scenario $\mathcal{S}1$  (blue band) both for $\mu$ and $\tau$ leptons. However, these values decreases when the magnitude of $S_{LL}$ decreases and it is clear from Fig. 4(a) and 4(c). The situation is similar when we plotted the $\mathcal{BR}_T$ as a function of $D_{LL}$ by fixing the parameters $\mathcal{B}_{sb},\phi_{sb},S_{LL}$ in the range given in Table IV, where one can see that it is also a decreasing function of $D_{LL}$. However, even for the small values of the $Z^{\prime}$ parameters, the value of the observable is quite distinct from the SM result especially in scenario $\mathcal{S}1$. Just like the Longitudinal polarized branching ratio, the effects of $Z^{\prime}$ boson corresponding to scenario $\mathcal{S}3$ (magenta bar in Fig. 4(c)) in the transverse polarized branching ratio are quite promising in $B_{s}\to \phi \tau^{+}\tau^{-}$ decay.
\end{itemize}

\begin{figure}[tbp]
\begin{tabular}{cc}
\epsfig{file=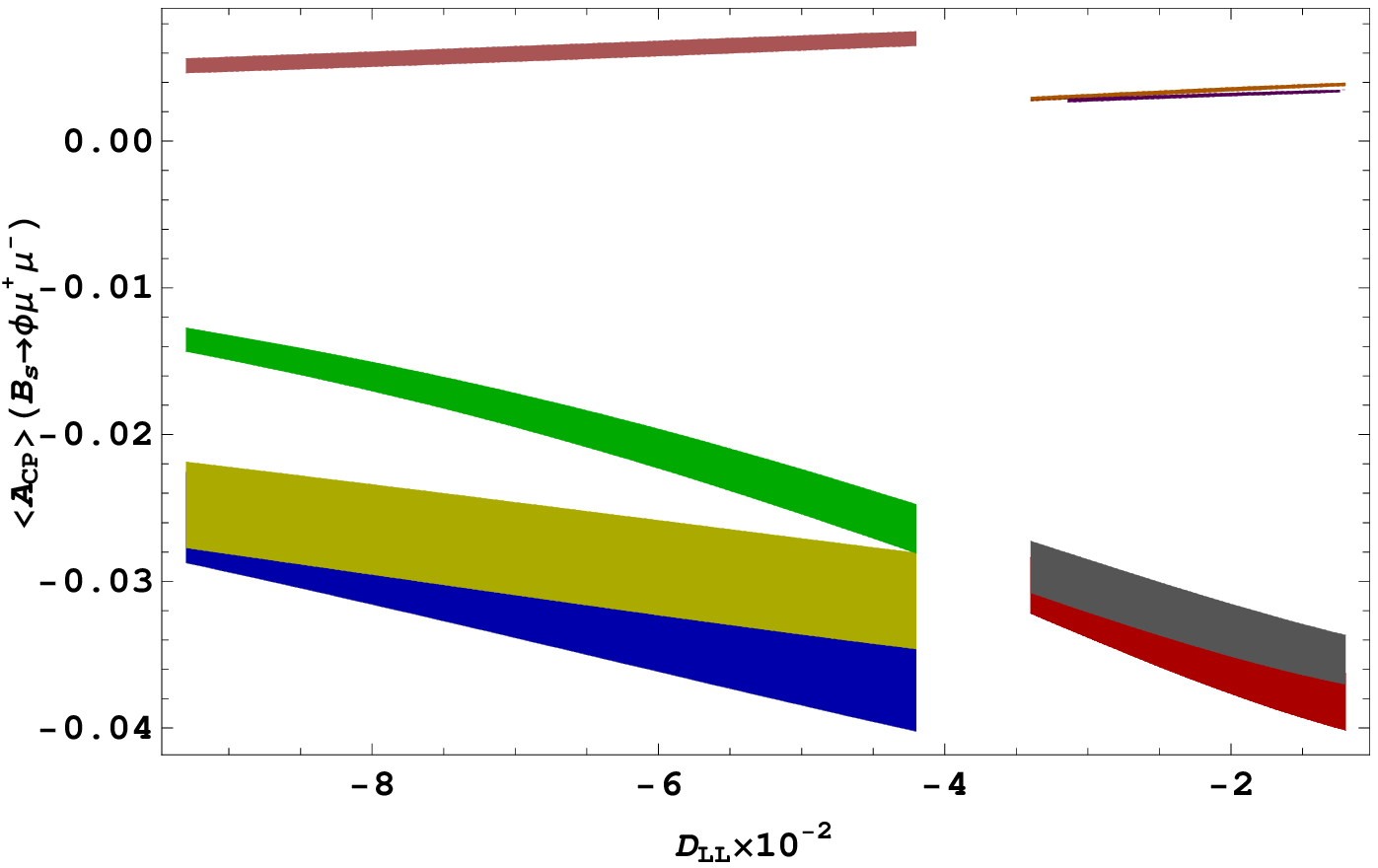,width=0.51\linewidth,clip=a} \put (-100,150){(a)} & %
\epsfig{file=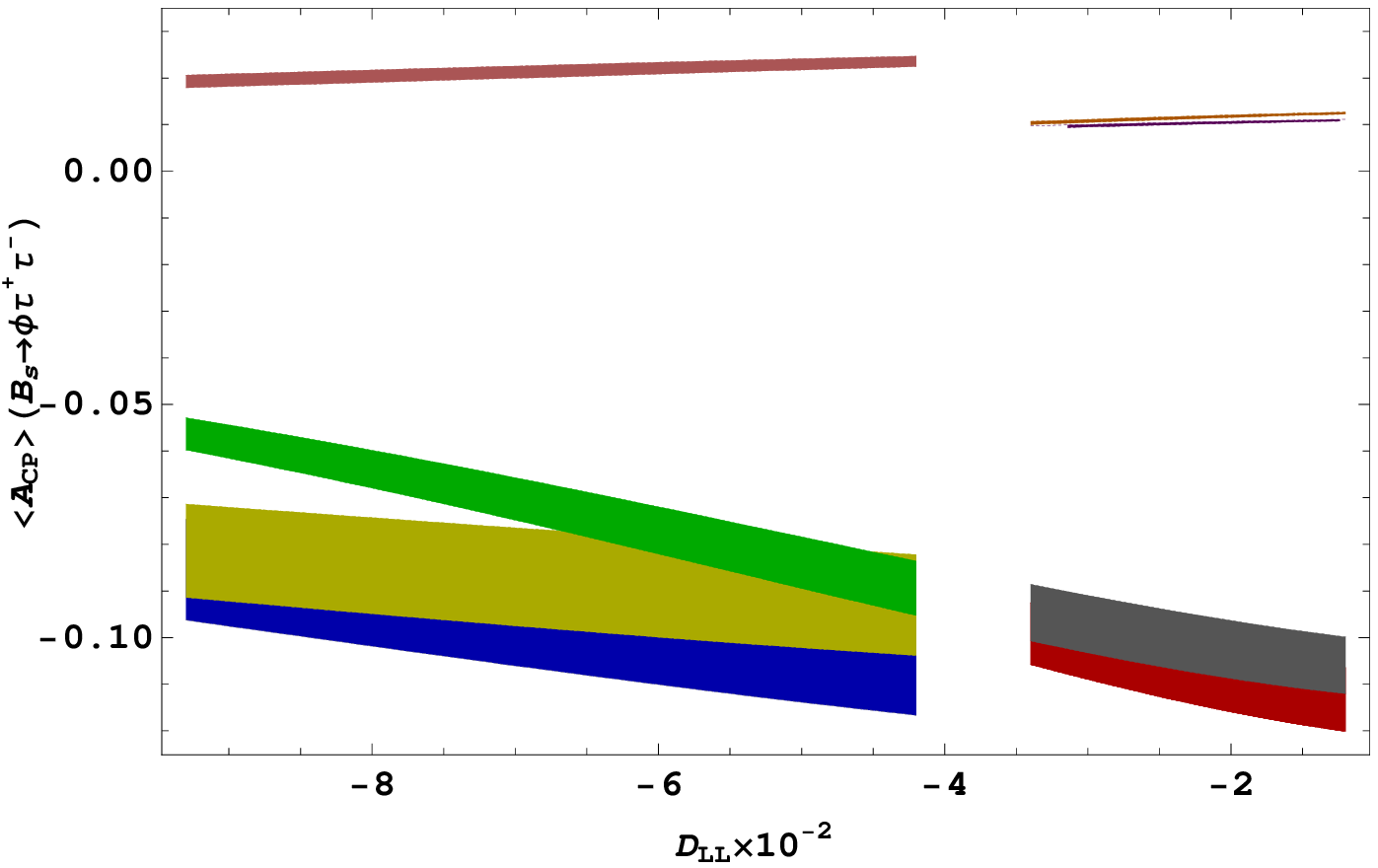,width=0.51\linewidth,clip=b} \put (-100,150){(b)}%
\end{tabular}%
\caption{Unpolarized $CP$ violation asymmetry $\mathcal{A}_{CP}$ as
function of $D_{LL}$ for $B_s\to \phi \mu^+ \mu^- (\tau^+\tau^-)$ for scenarios $\mathcal{S}1$ and
$\mathcal{S}2$. The Blue, Yellow, Green and Dark Pink colors correspond to the $\mathcal{S}1$ where as the
other colors are for the $\mathcal{S}2$. The band in each case depicts the variations of $\phi_{sb}$ in respective
scenario.} \label{UnCPmu}
\end{figure}

\begin{figure}[tbp]
\begin{tabular}{cc}
\epsfig{file=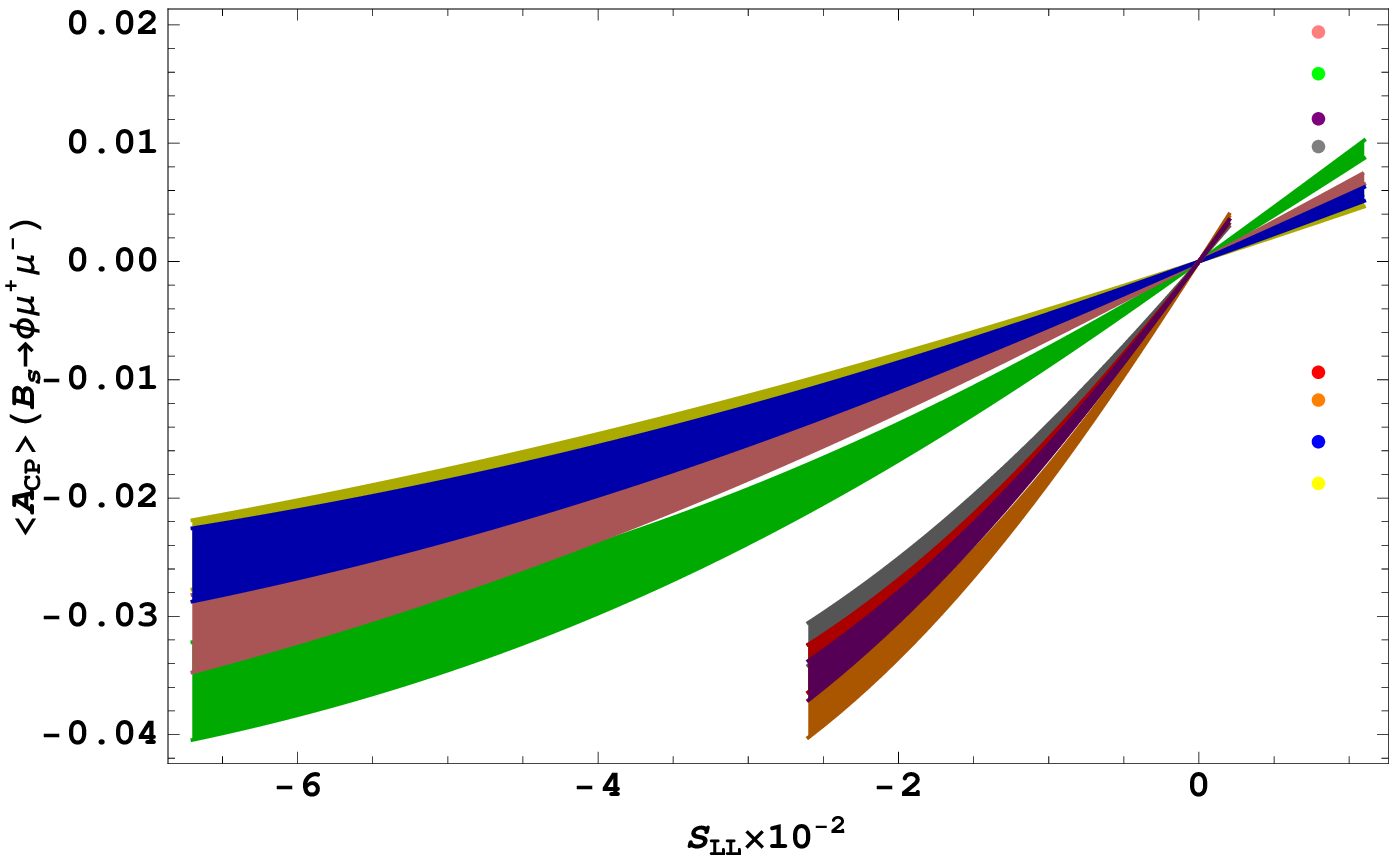,width=0.51\linewidth,clip=a} \put (-100,150){(a)} & %
\epsfig{file=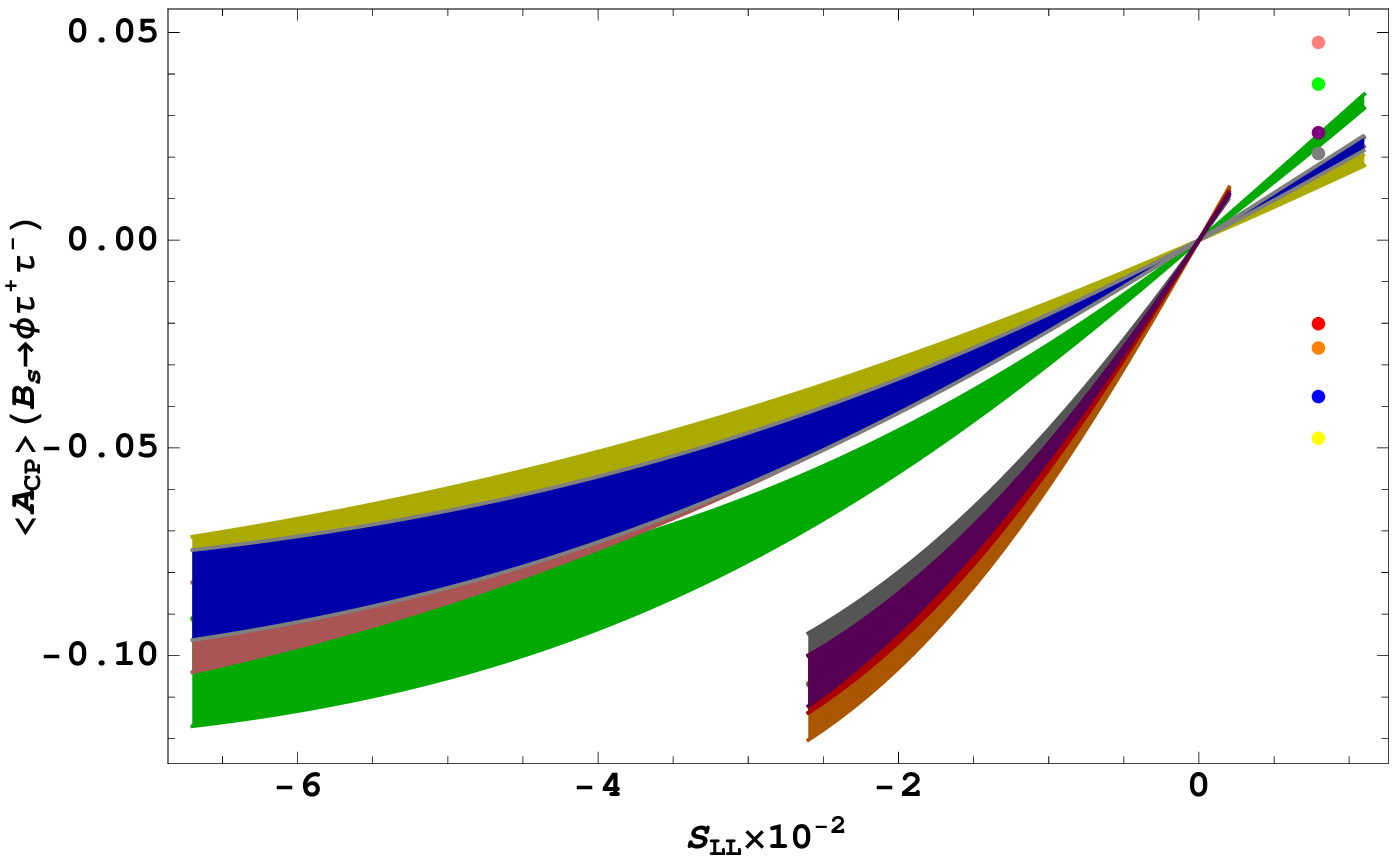,width=0.51\linewidth,clip=b} \put (-100,150){(b)}%
\end{tabular}%
\caption{Unpolarized $CP$ violation asymmetry $\mathcal{A}_{CP}$ as
function of $S_{LL}$ for $B_s\to \phi \mu^+ \mu^- (\tau^+\tau^-)$ for scenarios $\mathcal{S}1$, $\mathcal{S}2$ and
$\mathcal{S}3$. The color and band description is same as in Fig. \ref{UnCPmu}. The different color dots correspond to the
different values of $Z^{\prime}$ parameters in scenario $\mathcal{S}3$.} \label{UnCPmutau}
\end{figure}

\textbf{Unpolarized $CP$ violation asymmetry:}
\begin{itemize}
\item In Figs. \ref{UnCPmu} and \ref{UnCPmutau} the unpolarized $CP$ violation asymmetries for the $B_{s} \rightarrow \phi \mu^+ \mu^- (\tau^+ \tau^-)$ are
presented as a function of $D_{LL}$ and $S_{LL}$. It is well known that in SM the $CP$ violation asymmetry is almost zero, whereas,
by looking at the Eq. (\ref{unpolcp})
one can see that $\mathcal{A}_{CP}$ is proportional to the parameters of the $Z^{\prime}$ model which comes through
the imaginary part of the Wilson coefficients as well as of new weak phase $\phi_{sb}$ which is encoded in $\Lambda_{sb}$. Hence a significant non-zero
value gives us the clear indications of NP arising due to the extra neutral $Z^{\prime}$ boson. Therefore, we
are expecting a dependence on the new phase $\phi_{sb}$ and it is clear from Figs. \ref{UnCPmu} and \ref{UnCPmutau}
where band in each color depicts it. In Fig. \ref{UnCPmu} by changing the values of $S_{LL}, \phi_{sb}$ and $\mathcal{B}_{sb}$
the $\mathcal{A}_{CP}$ is plotted vs $D_{LL}$, where we can see that the value
is not appreciably changed when we have muon as final state leptons. However, in case of the tau leptons (c.f. Fig. \ref{UnCPmu}(b)), the value of $\mathcal{A}_{CP}$ is around $-0.11$ in scenarios $\mathcal{S}1(\mathcal{S}2)$ for the $D_{LL}=-4.1\times 10^{-2}(-1.6\times 10^{-2})$ and $S_{LL}=-6.7\times 10^{-2}(-2.6\times 10^{-2})$ shown by blue (red) bands.

\item Fig. \ref{UnCPmutau} presents the behavior of $\mathcal{A}_{CP}$ with $S_{LL}$ by varying the values of $D_{LL},
\phi_{sb}$ and $\mathcal{B}_{sb}$ in the range given in Table IV. Again it can be seen that in case of muon the value is small compared
to the case when $\tau$'s are final state leptons. In both cases $\mathcal{A}_{CP}$ is an increasing function of the $S_{LL}$.
In $B_{s} \to \phi \tau^+ \tau^-$ the value of unpolarized $CP$ asymmetry is around $-0.12$ for certain values of $Z^{\prime}$
parameters in both scenarios $\mathcal{S}1$ and $\mathcal{S}2$.

The values of unpolarized $CP$ violation asymmetry in scenario $\mathcal{S}3$ for $B_{s}\to \phi \mu^{+}\mu^{-}$ and $B_{s}\to \phi \tau^{+}\tau^{-}$ are shown by different color dots in Figs. 6(a) and 6(b), respectively. It can be noticed that the value of unpolarized $CP$ violation asymmetry is maximum in this scenario when $\phi_{sb}=160^{\circ}$, $|\mathcal{B}_{sb}|=5\times 10^{-3}$ and it is depicted by the orange dot in these figures. When the new weak phase $(\phi_{sb})$ has the negative value, the value of the unpolarized $CP$ violation asymmetry is just opposite to the case when $\phi_{sb}$ is positive. This is shown by the lower four dots in the Figs. 6(a) and 6(b).
\end{itemize}

\begin{figure}[tbp]
\begin{tabular}{cc}
\epsfig{file=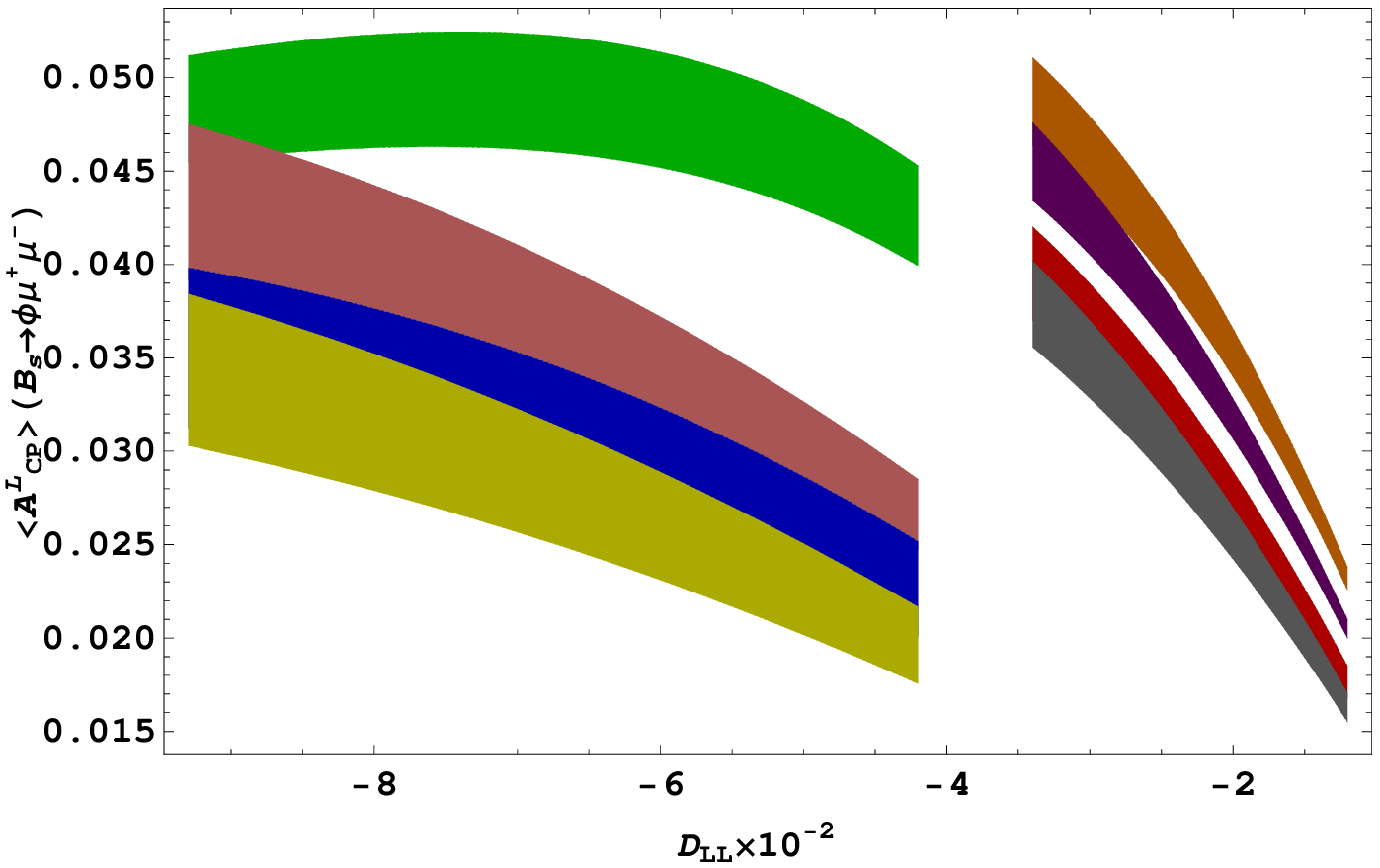,width=0.51\linewidth,clip=a} \put (-100,150){(a)} & %
\epsfig{file=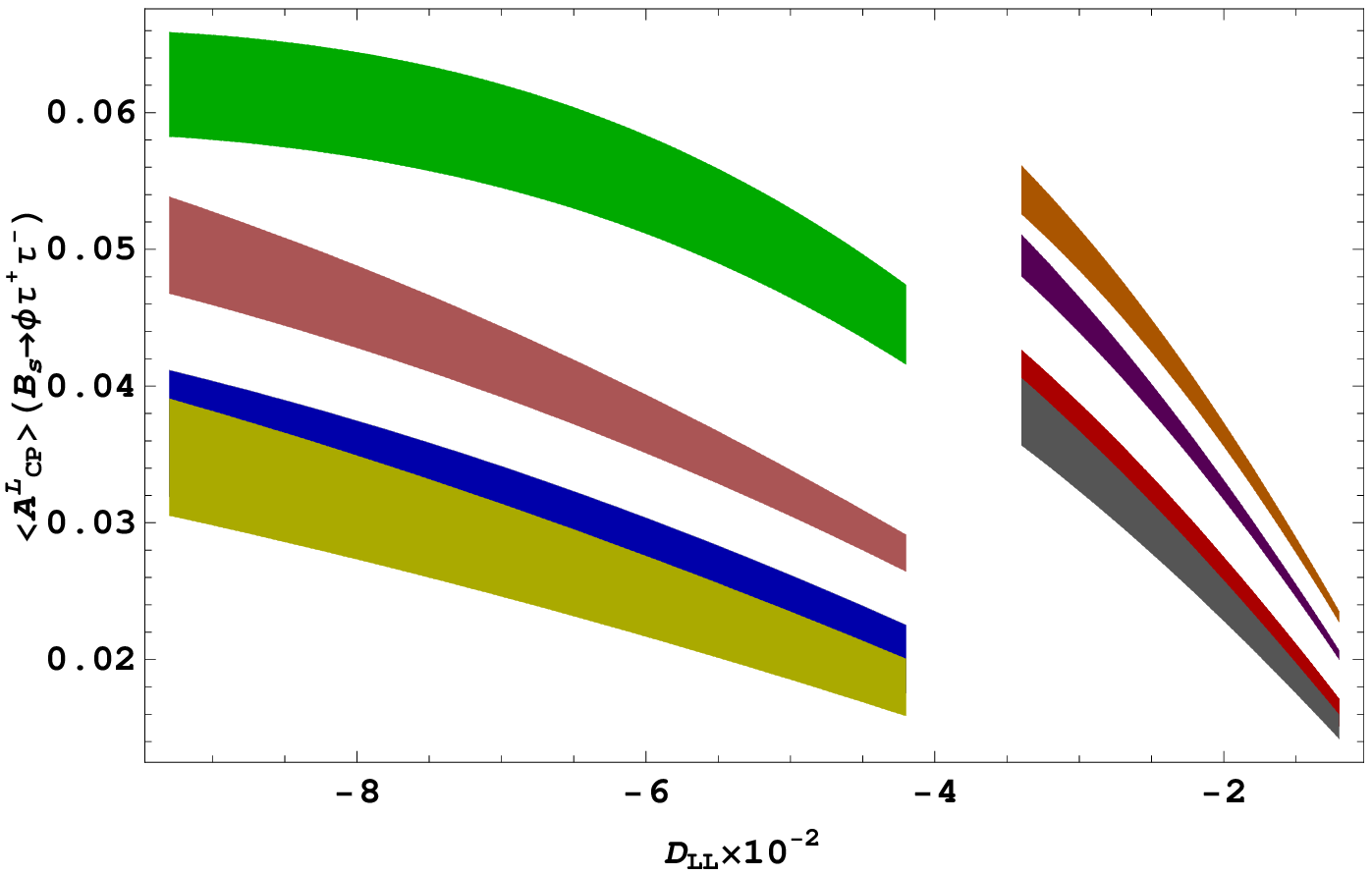,width=0.51\linewidth,clip=b} \put (-100,150){(b)}%
\end{tabular}%
\caption{Longitudinal polarized $CP$ violation asymmetry $\mathcal{A}^{L}_{CP}$ as
function of $D_{LL}$ for $B_s\to \phi \mu^+ \mu^- (\tau^+\tau^-)$ for scenarios $\mathcal{S}1$ and
$\mathcal{S}2$. The color and band description is same as in Fig. \ref{UnCPmu}.} \label{LnCPmu}
\end{figure}

\begin{figure}[tbp]
\begin{tabular}{cc}
\epsfig{file=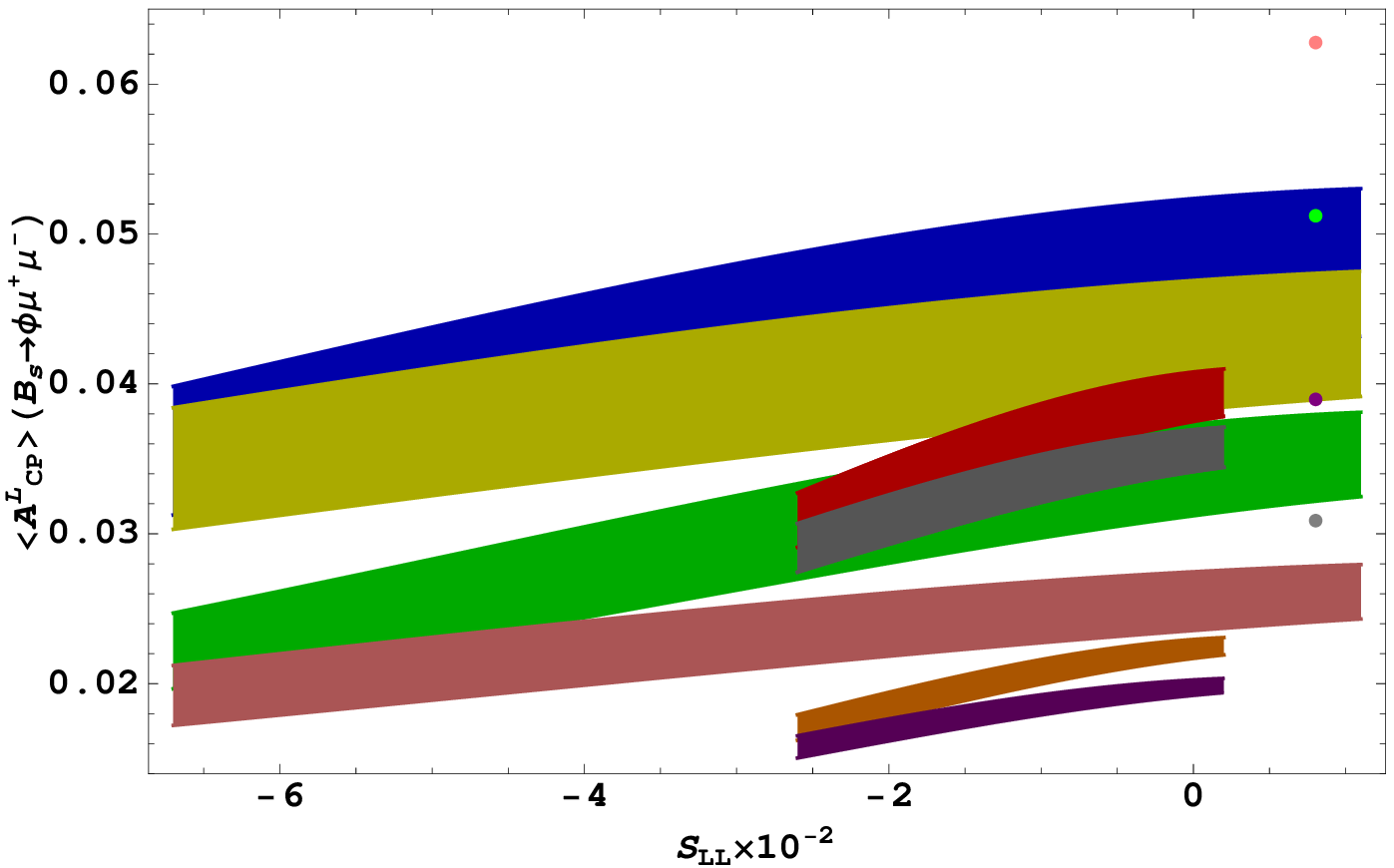,width=0.51\linewidth,clip=a} \put (-100,150){(a)} & %
\epsfig{file=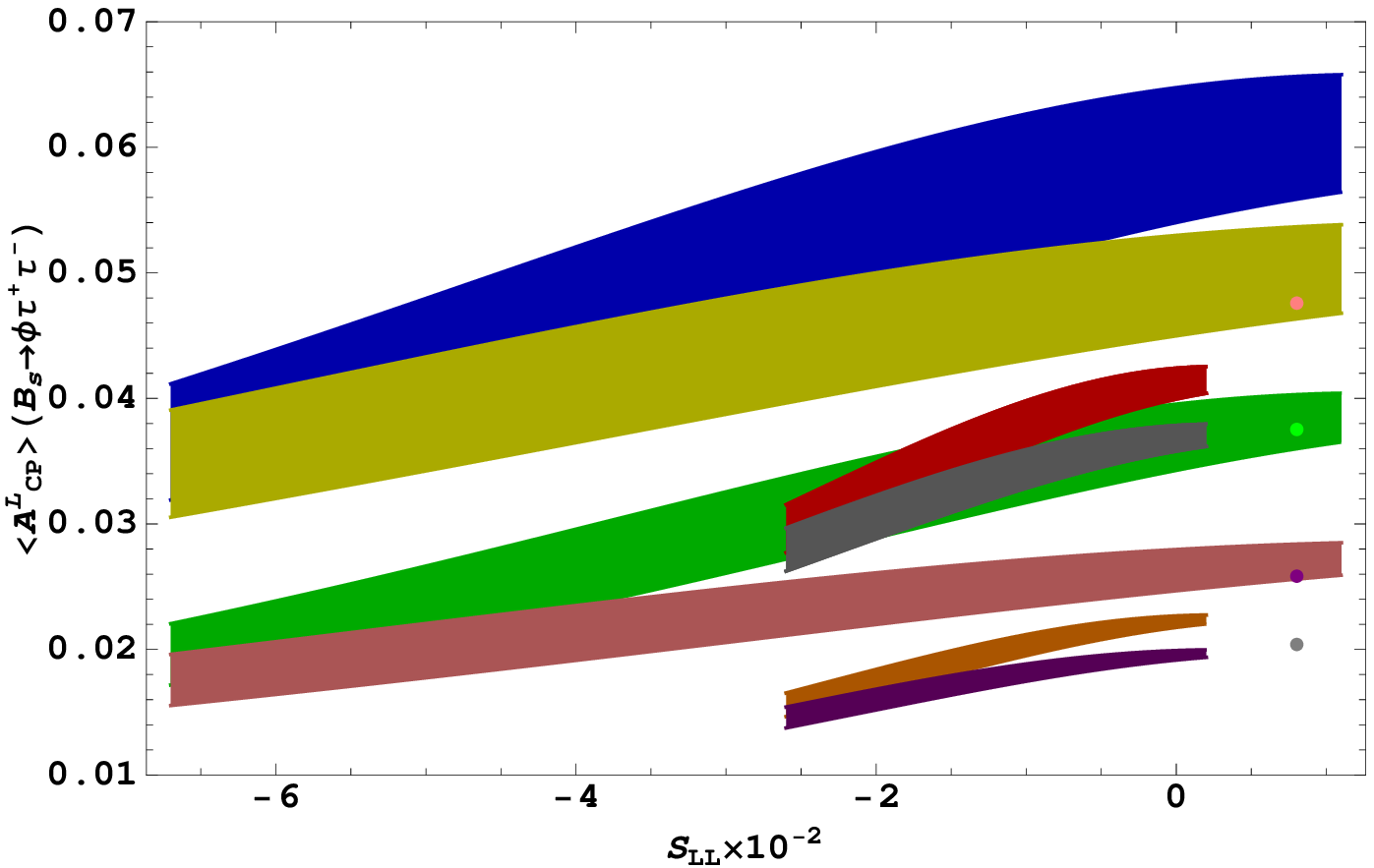,width=0.51\linewidth,clip=b} \put (-100,150){(b)}%
\end{tabular}%
\caption{Longitudinal polarized $CP$ violation asymmetry $\mathcal{A}^{L}_{CP}$ as
function of $S_{LL}$ for $B_s\to \phi \mu^+ \mu^- (\tau^+\tau^-)$ for scenarios $\mathcal{S}1$ and
$\mathcal{S}2$. The color and band description is same as in Fig. \ref{UnCPmu}. The different color dots correspond to the
different values of $Z^{\prime}$ parameters in scenario $\mathcal{S}3$.} \label{LnCPtau}
\end{figure}

\textbf{Longitudinal polarized $CP$ violation asymmetry:}
\begin{itemize}
\item The longitudinal polarized $CP$ violation asymmetry $\mathcal{A}^L_{CP}$ is drawn
in Figs. \ref{LnCPmu} and \ref{LnCPtau}. From Eq. (\ref{Qfunc}) it can be noticed that $\mathcal{Q}^{L}$ is proportional to the
imaginary part of the combination of Wilson coefficients $C_{9}$ and $C_{10}$ both in the SM as well as in the $Z^{\prime}$ model. This makes
the $\mathcal{A}^L_{CP}$ sensitive to the change in the values of these Wilson coefficients in
the $Z^{\prime}$ model. In Fig. \ref{LnCPmu}(a) and \ref{LnCPmu}(b), we have plotted the $\mathcal{A}^L_{CP}$ vs $D_{LL}$ by fixing the values of $S_{LL}$ and other $Z^{\prime}$ parameters in the range given in Table IV. We can see that the value of $\mathcal{A}^L_{CP}$
increases from $0.015$ to $0.055$ when muons are the final state leptons and from $0.018$ to $0.068$
in case of tau's as final state leptons which can be visualized from the green(orange) band that corresponds to scenario $\mathcal{S}1(\mathcal{S}2)$.
The situation when the longitudinal polarized $CP$ violation asymmetry
is plotted with $S_{LL}$ by taking other parameters in the range given in Table IV and it is displayed in Fig.\ref{LnCPtau}. Here we can see that it is an increasing function of $S_{LL}$ where in $\mathcal{S}1$ the value increase from $0.040(0.042)$ to $0.052(0.064)$ when we have $\mu^{+}\mu^{-}(\tau^{+}\tau^{-})$ final state leptons and it is clearly visible from the blue band. In comparison, for $\mathcal{S}2$
these values increase from $0.028$ to $0.036$ both in case of $\mu$ and $\tau$ leptons. It can also be seen in Fig. 8, the value of longitudinal polarized $CP$ violation asymmetry in scenario $\mathcal{S}3$ lies in the ball park of first two scenarios except the limit when $\phi _{sb}=160^{\circ}$, $|\mathcal{B}_{sb}|=5\times 10^{-3}$. For this value, one can see that the value of longitudinal polarized $CP$ violation asymmetry in
 $B_{s}\to \phi \mu^{+}\mu^{-}$ is around $0.061$ which is significantly different from its value in $\mathcal{S}1$ and $\mathcal{S}2$. Hence, by measuring
$\mathcal{A}^L_{CP}$ one can not only segregate the NP coming through the $Z^{\prime}$ boson but can also distinguish the
three scenarios, named here as, $\mathcal{S}1$, $\mathcal{S}2$ and $\mathcal{S}3$.
\end{itemize}
\begin{figure}[tbp]
\begin{tabular}{cc}
\epsfig{file=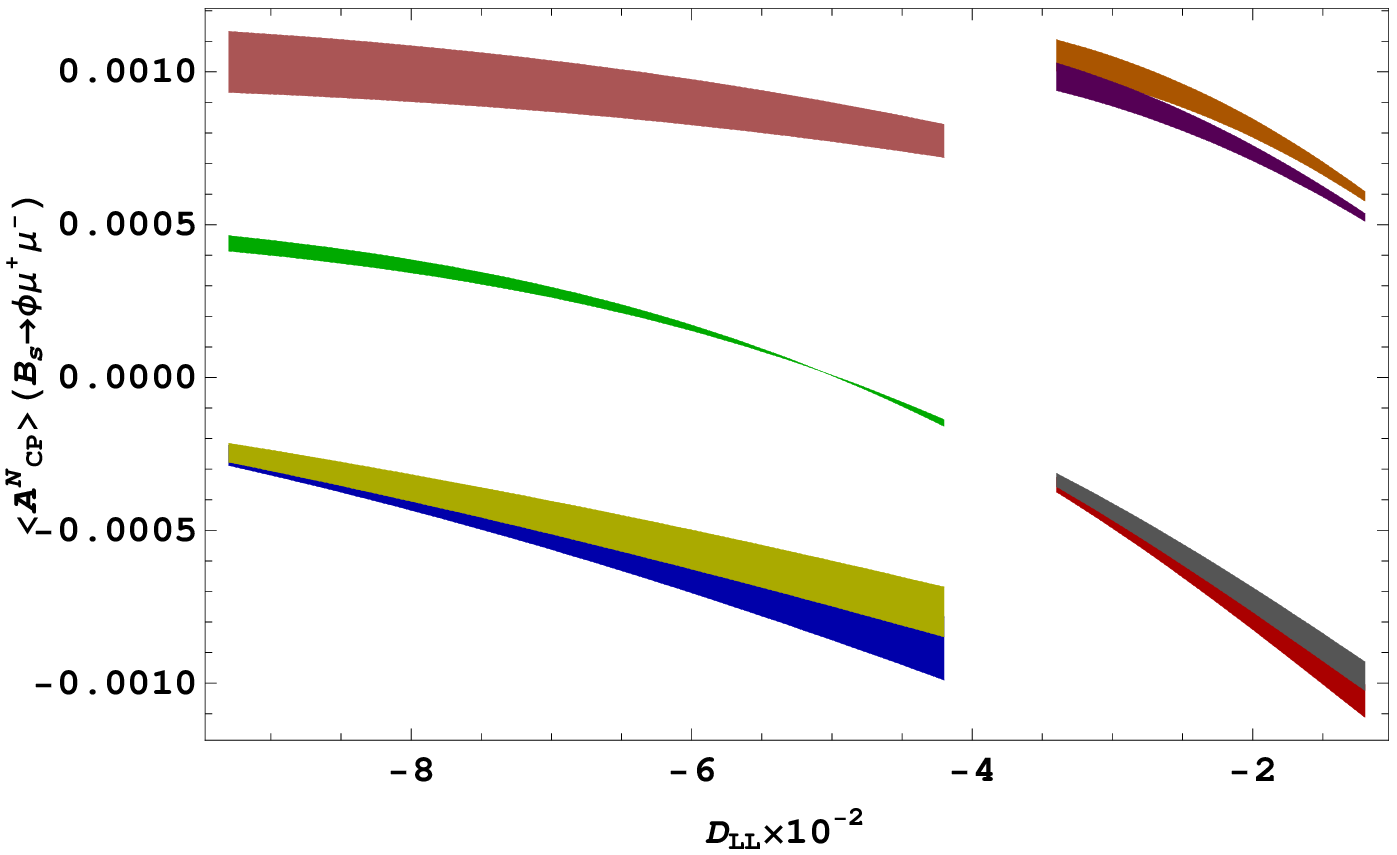,width=0.51\linewidth,clip=a} \put (-100,150){(a)} & %
\epsfig{file=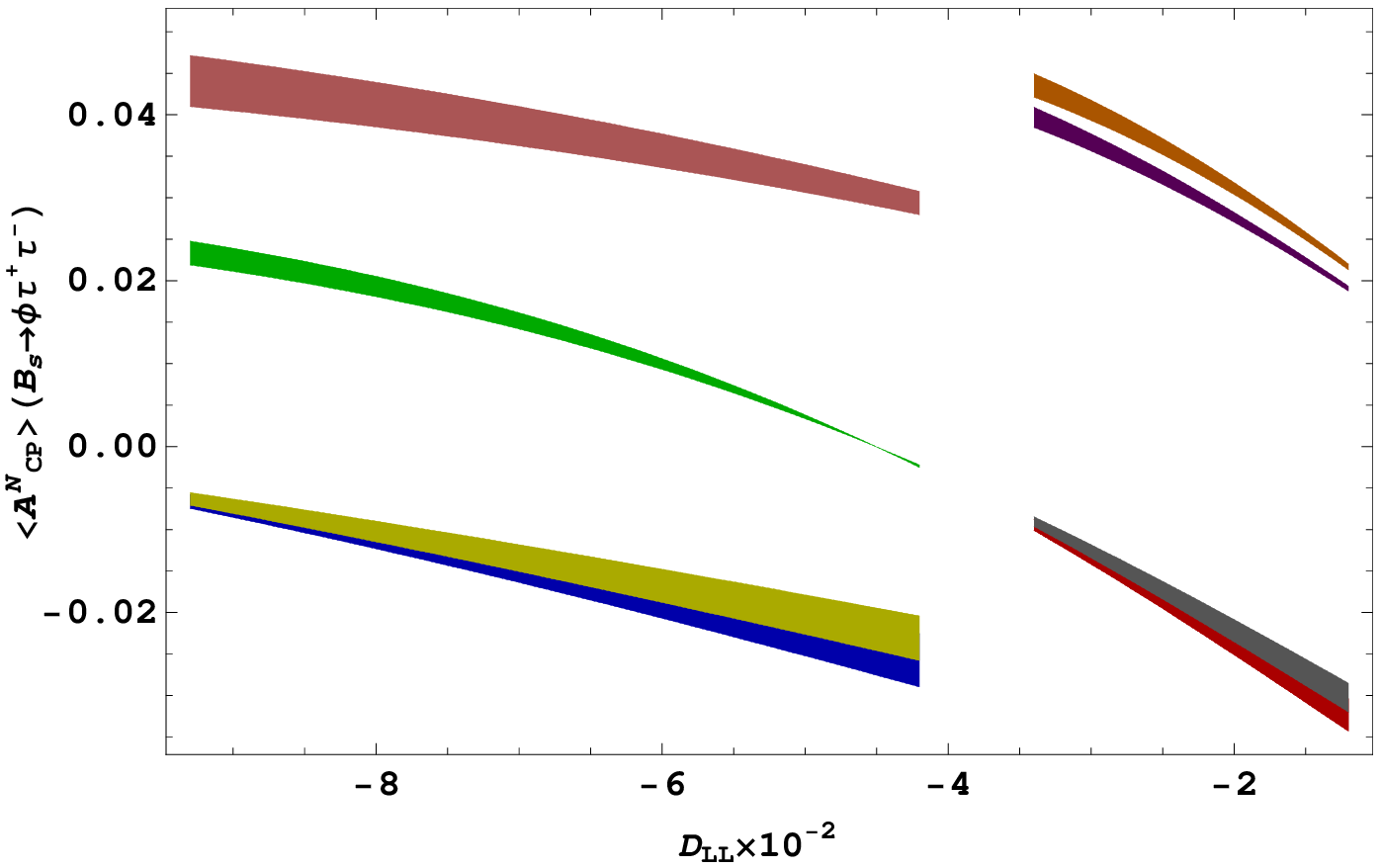,width=0.51\linewidth,clip=b} \put (-100,150){(b)}%
\end{tabular}%
\caption{Normal polarized $CP$ violation asymmetry $\mathcal{A}^{N}_{CP}$ as
function of $D_{LL}$ for $B_s\to \phi \mu^+ \mu^- (\tau^+\tau^-)$ for scenarios $\mathcal{S}1$ and
$\mathcal{S}2$. The color and band description is same as in Fig. \ref{UnCPmu}.} \label{NoCPmu}
\end{figure}

\begin{figure}[tbp]
\begin{tabular}{cc}
\epsfig{file=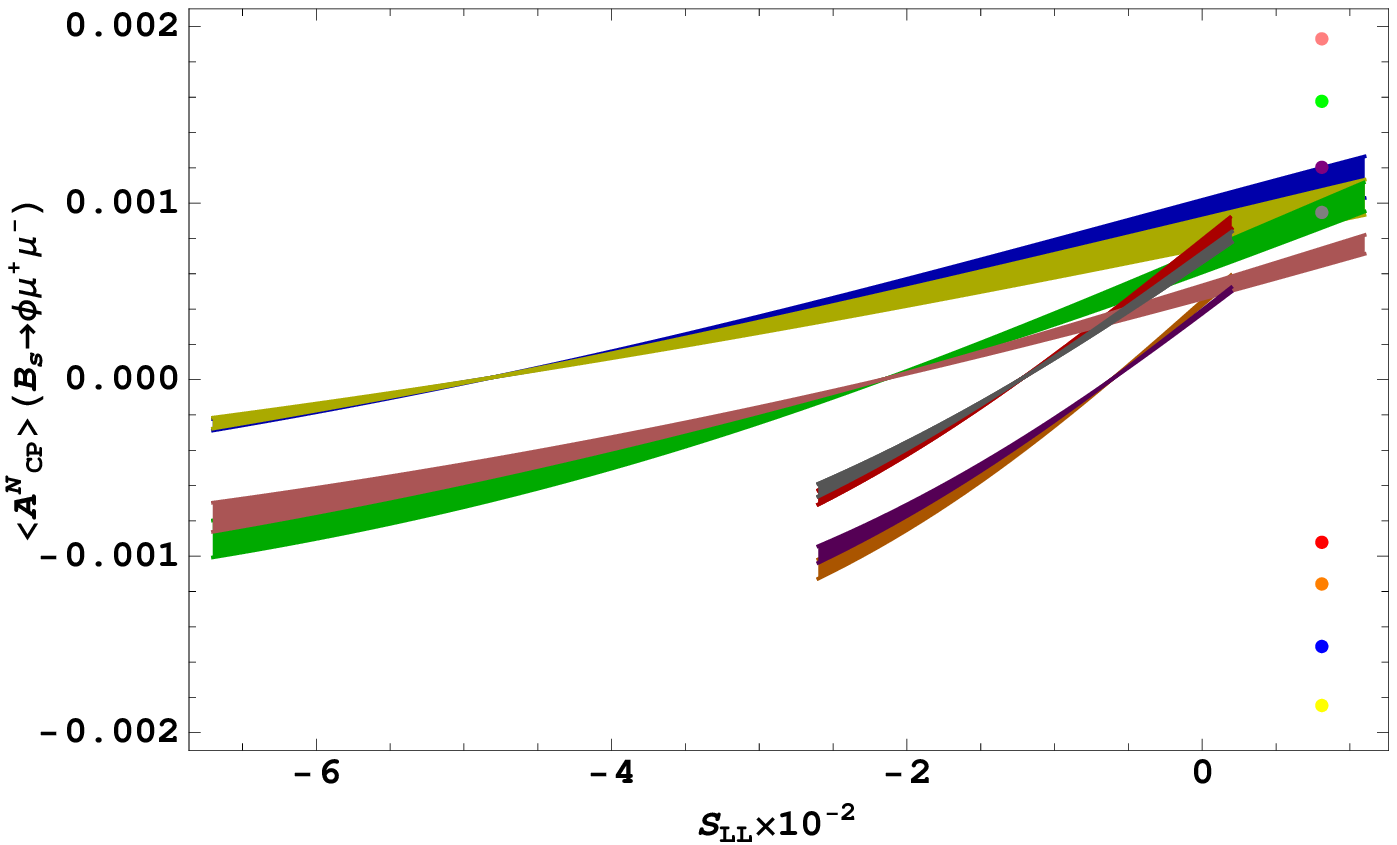,width=0.51\linewidth,clip=a} \put (-100,150){(a)} & %
\epsfig{file=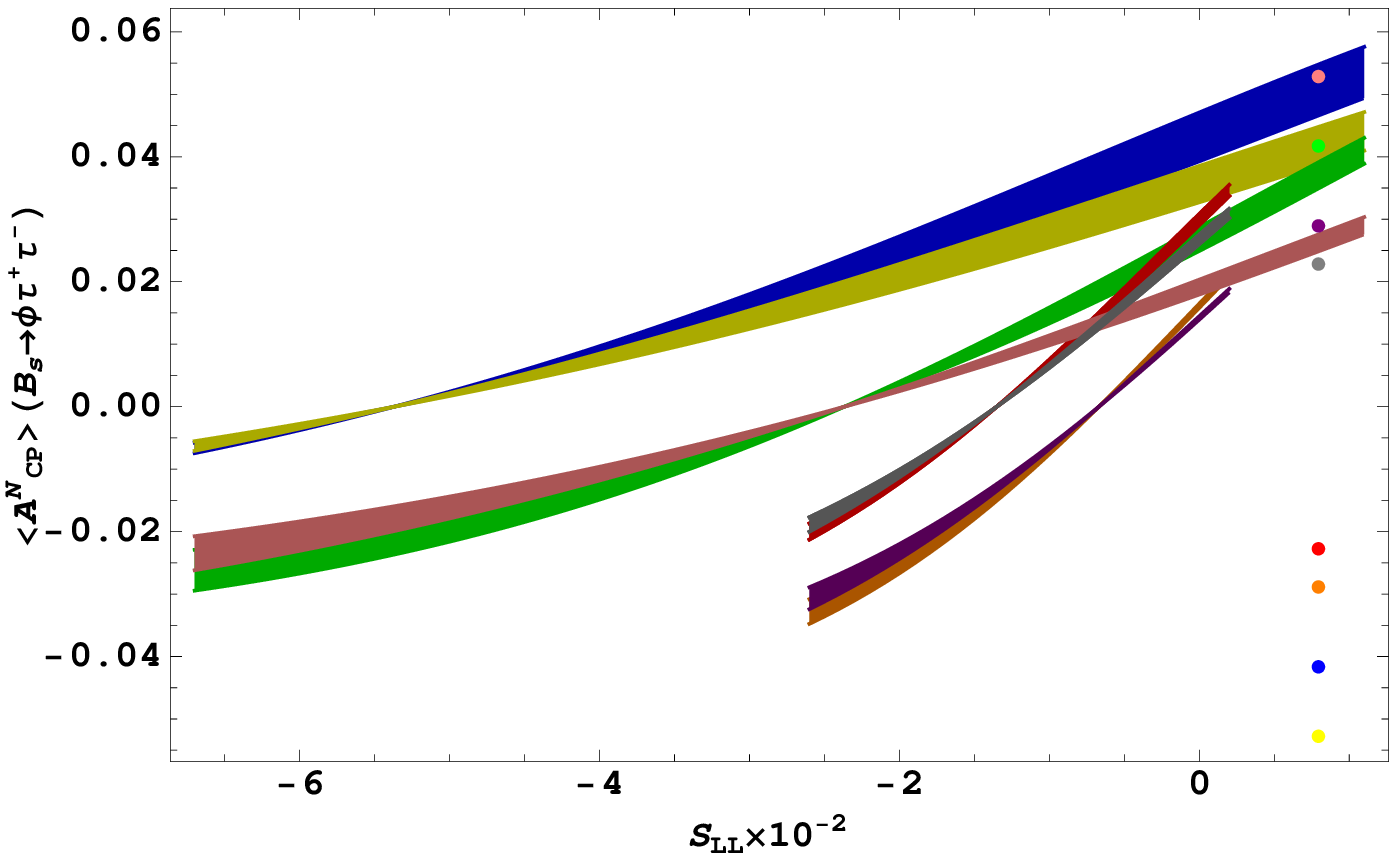,width=0.51\linewidth,clip=b} \put (-100,150){(b)}%
\end{tabular}%
\caption{Normal polarized $CP$ violation asymmetry $\mathcal{A}^{N}_{CP}$ as
function of $S_{LL}$ for $B_s\to \phi \mu^+ \mu^- (\tau^+\tau^-)$ for scenarios $\mathcal{S}1$ and
$\mathcal{S}2$. The color and band description is same as in Fig. \ref{UnCPmu}. The different color dots correspond to the
different values of $Z^{\prime}$ parameters in scenario $\mathcal{S}3$.} \label{NoCPtau}
\end{figure}

\textbf{Normal polarized $CP$ violation asymmetry:}
\begin{itemize}
\item In contrast to the $\mathcal{A}_{CP}$ and $\mathcal{A}^{L}_{CP}$, the
normal polarized $CP$ violation asymmetry is an order of magnitude smaller in case of muon compared to the tauon as final state leptons.
Let us try to understand it from the expressions presented in Eq. (\ref{Qfunc}). The $\mathcal{A}^{N}_{CP}$ comes from the function
$\mathcal{Q}^{N}$ which contains $\mathcal{H}_{4}, \mathcal{H}_5$ and $\mathcal{H}_6$. In Eq. (\ref{Hfunctions}) it is clear
that these are proportional to the lepton mass and their suppression in case of muon is obvious and Figs. \ref{NoCPmu}(a) and
\ref{NoCPtau}(a) depicts this fact.
Coming to the Figs. \ref{NoCPmu}(b) and \ref{NoCPtau}(b) we can see that the
$\mathcal{A}^{N}_{CP}$ is very sensitive to the parameters of $Z^{\prime}$ both in
the $\mathcal{S}1$ and $\mathcal{S}2$ where similar to the $\mathcal{A}_{CP}$ it changes its sign.
In Fig. \ref{NoCPmu}(b), the value of $\mathcal{A}^{N}_{CP}$ decreases from $0.042$ to $-0.018$ in the parameter range of $Z^{\prime}$
in $\mathcal{S}1$ and from $0.043$ to $-0.014$ in $\mathcal{S}2$.
In contrast Fig. \ref{NoCPtau}(b) depicts the case where $\mathcal{A}^{N}_{CP}$ is plotted with $S_{LL}$. Here we can see
that the value of $\mathcal{A}^{N}_{CP}$ increases from $-0.018$ to $0.055$ in $\mathcal{S}1$ and $-0.035$ to $0.035$ in the
second scenario $\mathcal{S}2$. In scenario $\mathcal{S}3$ the maximum value of normal $CP$ violation asymmetry is $0.05$ when we have $\tau^{+}\tau^{-}$ as final state leptons and the values of $\phi_{sb}=160^{\circ}$ and $|\mathcal{B}_{sb}|=5\times 10^{-3}$ . It is shown in Fig. 10 with the orange dot.
\end{itemize}
\begin{figure}[tbp]
\begin{tabular}{cc}
\epsfig{file=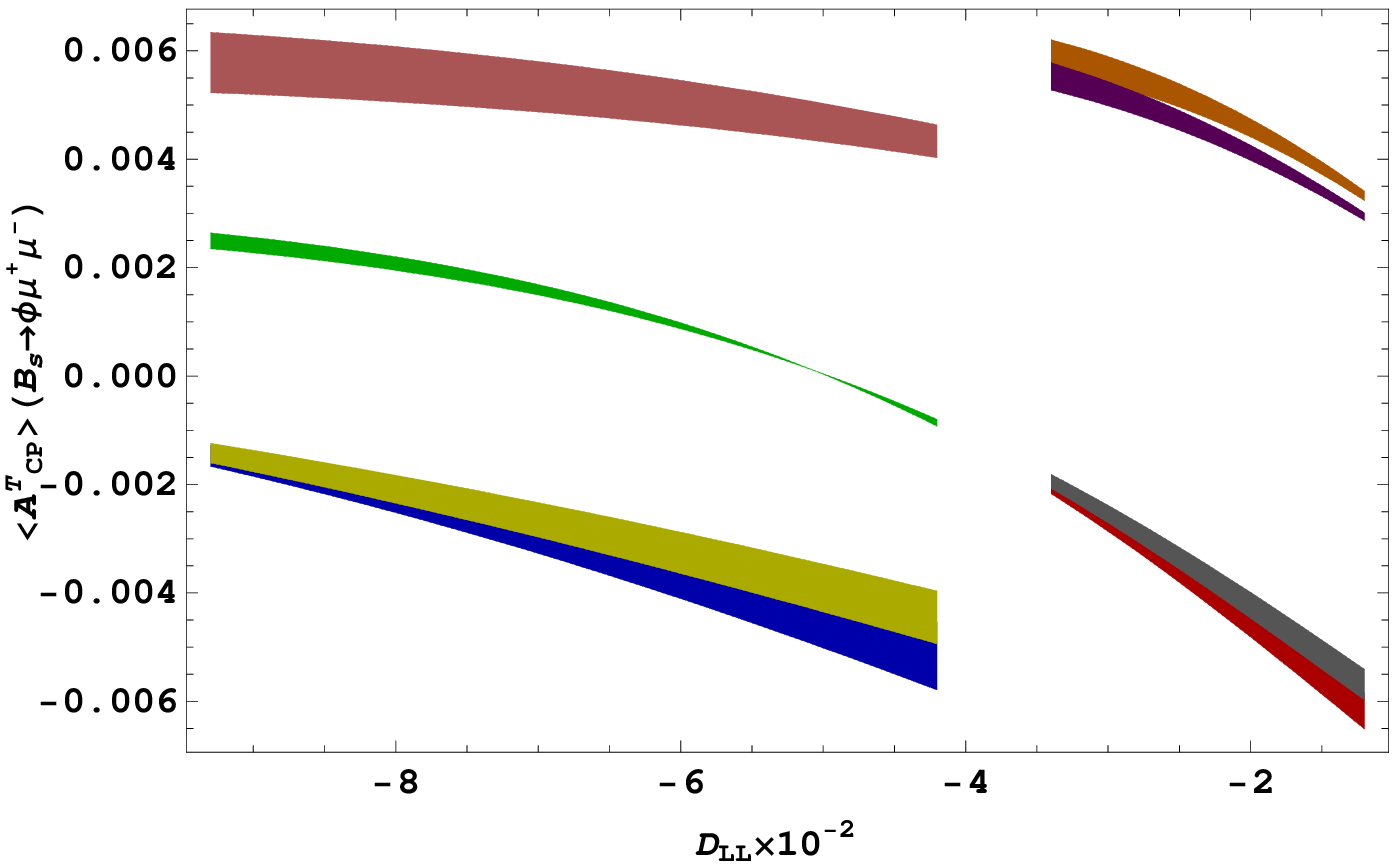,width=0.51\linewidth,clip=a} \put (-100,150){(a)} & %
\epsfig{file=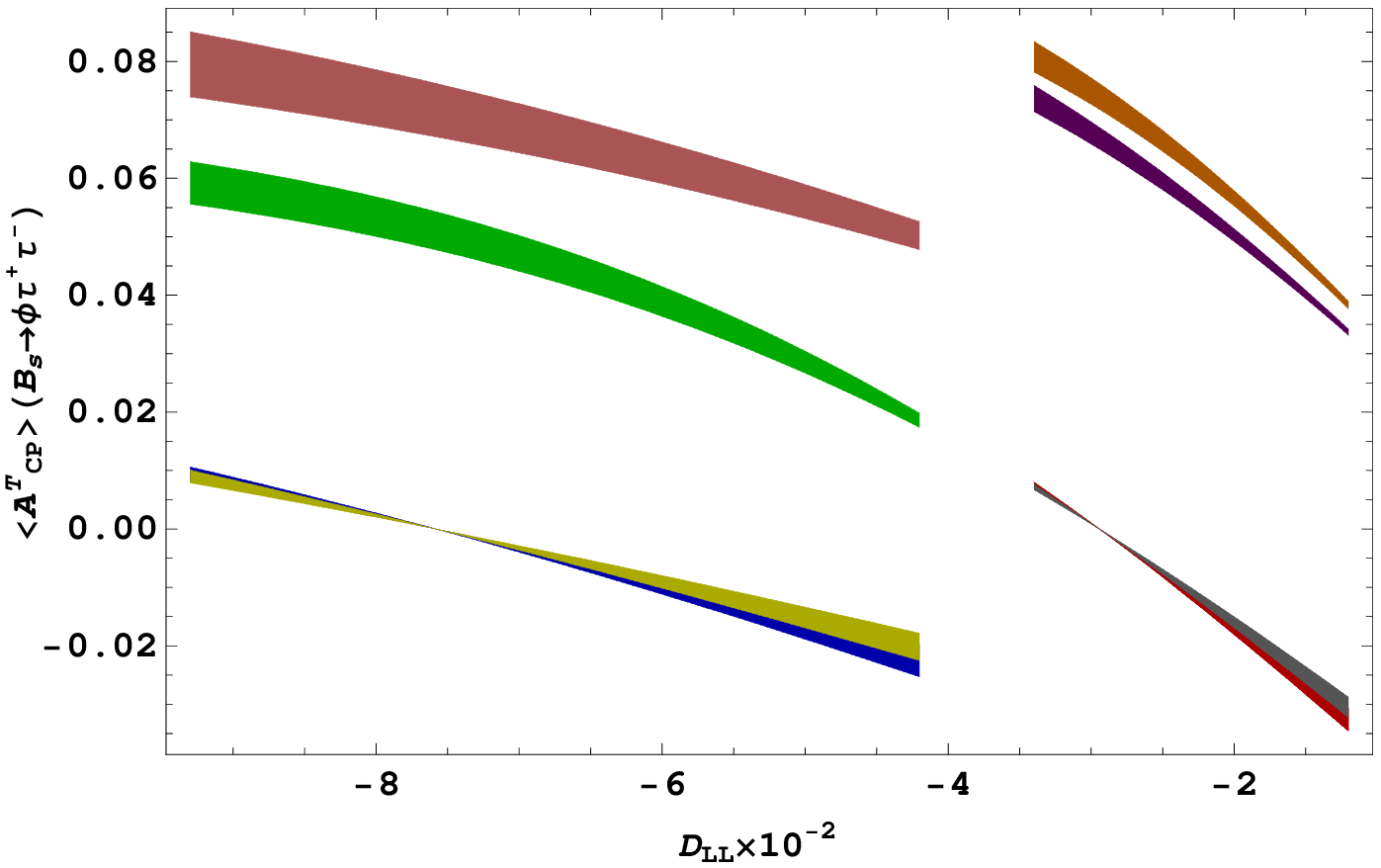,width=0.51\linewidth,clip=b} \put (-100,150){(b)}%
\end{tabular}%
\caption{Transversally polarized $CP$ violation asymmetry $\mathcal{A}^{N}_{CP}$ as
function of $D_{LL}$ for $B_s\to \phi \mu^+ \mu^- (\tau^+\tau^-)$ for scenarios $\mathcal{S}1$ and
$\mathcal{S}2$. The color and band description is same as in Fig. \ref{UnCPmu}.} \label{TrCPmu}
\end{figure}

\begin{figure}[tbp]
\begin{tabular}{cc}
\epsfig{file=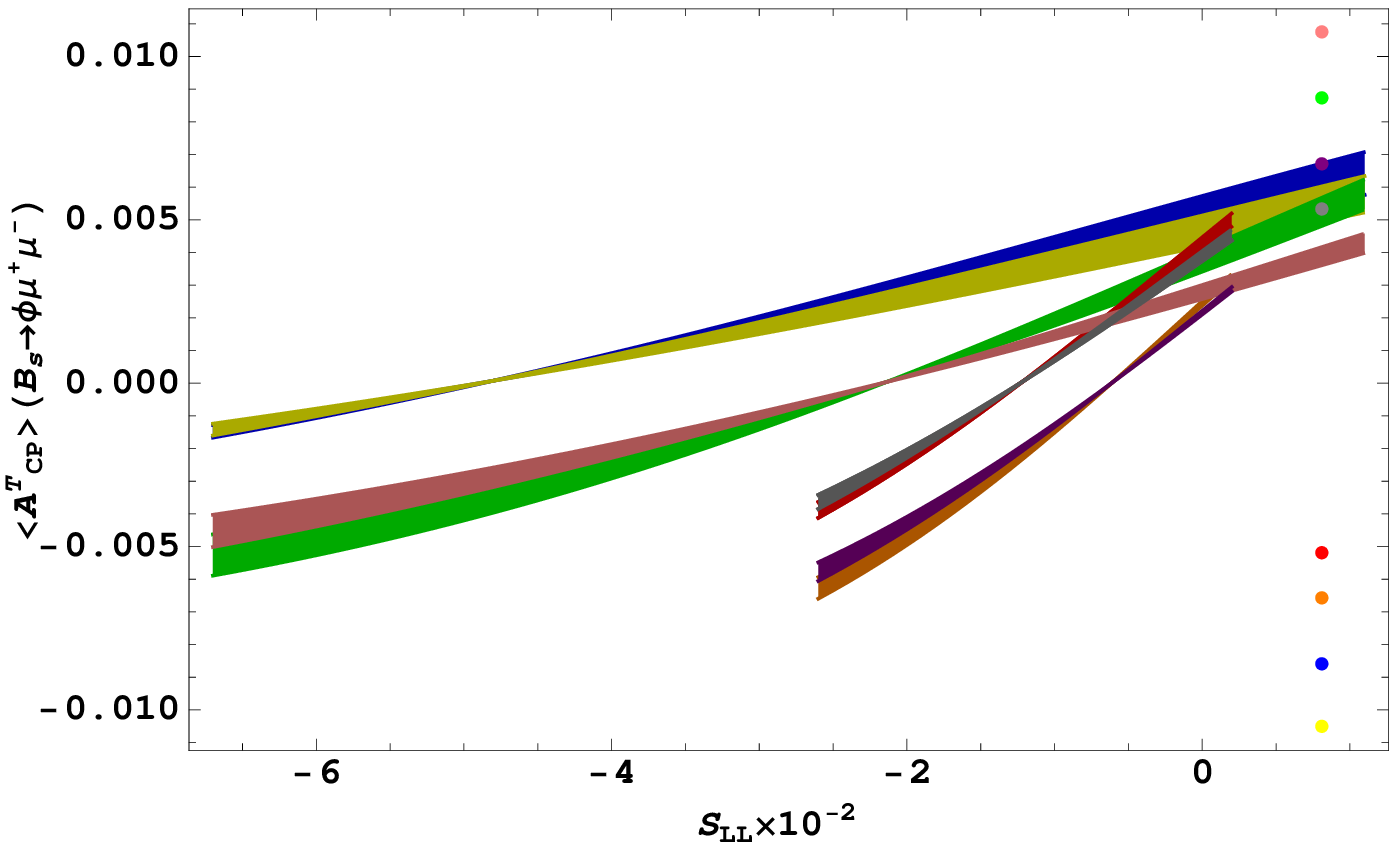,width=0.51\linewidth,clip=a} \put (-100,150){(a)} & %
\epsfig{file=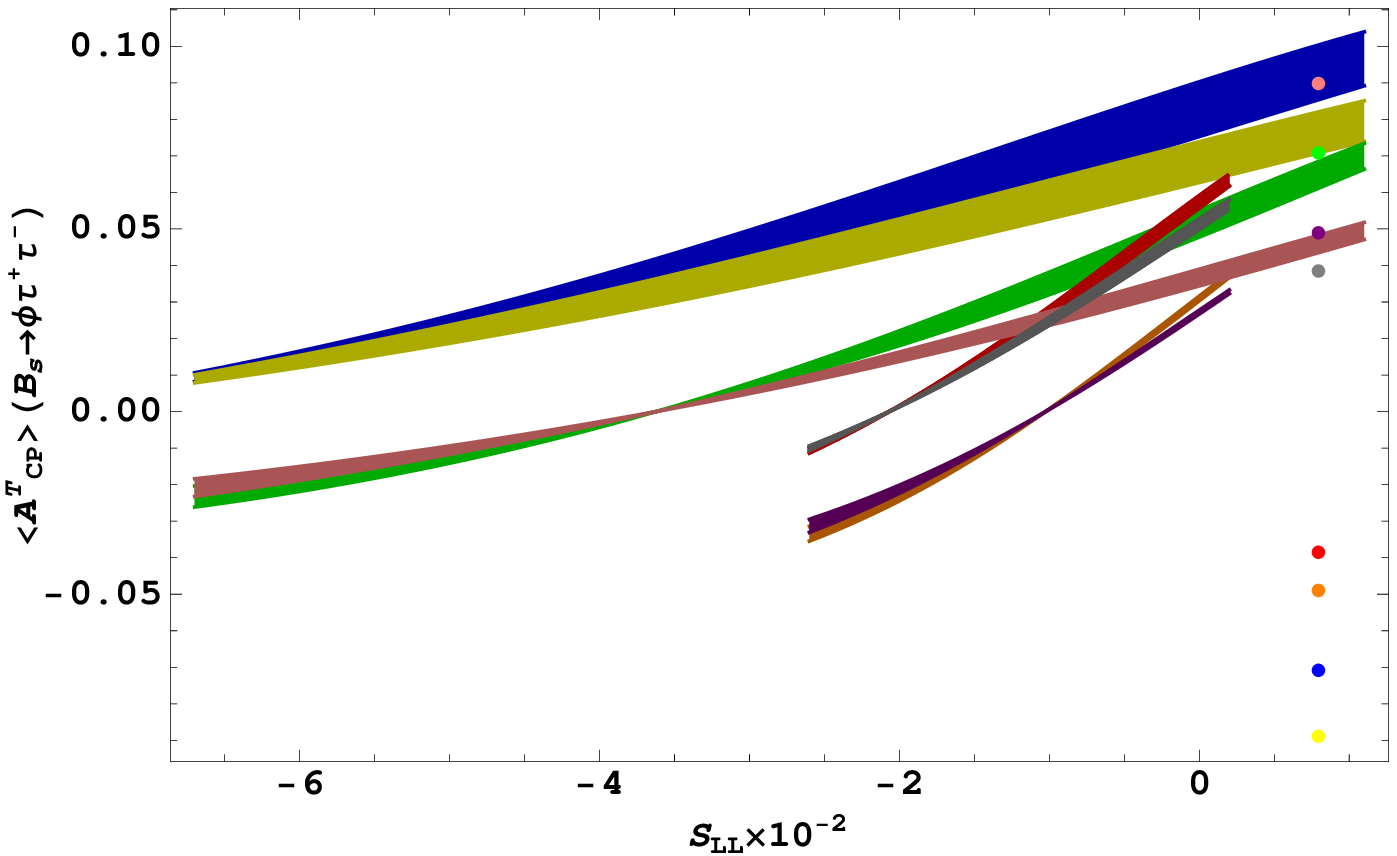,width=0.51\linewidth,clip=b} \put (-100,150){(b)}%
\end{tabular}%
\caption{Transversally polarized $CP$ violation asymmetry $\mathcal{A}^{N}_{CP}$ as
function of $S_{LL}$ for $B_s\to \phi \mu^+ \mu^- (\tau^+\tau^-)$ for scenarios $\mathcal{S}1$ and
$\mathcal{S}2$. The color and band description is same as in Fig. \ref{UnCPmu}. The different color dots correspond to the
different values of $Z^{\prime}$ parameters in scenario $\mathcal{S}3$.} \label{TrCPtau}
\end{figure}

\textbf{Transverse polarized $CP$ violation asymmetry:}
\begin{itemize}
\item In a same fashion, the
transverse polarized $CP$ violation asymmetry $\mathcal{A}^{T}_{CP}$ is also $m_{l}$ suppressed which is
visible from $\mathcal{H}_{6}$ appearing in the function $\mathcal{Q}^T$. The graphs given in Figs. \ref{TrCPmu}(a) and \ref{TrCPtau}(a)
just strengthen this argument, where $\mathcal{A}^{T}_{CP}$ is an order of magnitude suppressed in $B_s \rightarrow \phi \mu^+ \mu^-$
compared to the $B_s \rightarrow \phi \tau^+ \tau^-$. From Figs. \ref{TrCPmu}(b) and \ref{TrCPtau}(b)
it is clear that in case of the $\tau$'s as final state
leptons the value of the $\mathcal{A}^{T}_{CP}$ reaches to $0.1$ in certain parametric space of the $Z^{\prime}$ scenario $\mathcal{S}1$.

By varying the $Z^{\prime}$ parameters in the range given in Eq. (\ref{Newcons}) the trend of transverse $CP$ violation asymmetry is shown by different colors of dots in Fig. 12. For $\phi_{sb}=160^{\circ}, |\mathcal{B}_{sb}|=5\times 10^{-3}$ in scenario $\mathcal{S}3$ the value of transverse polarized $CP$ violation asymmetry in $B_{s}\to \phi \tau^{+}\tau^{-}$ is close to its maximum value in $\mathcal{S}1$ and it is shown by the orange dot in the Fig. 12(b).  This can be measured in different colliders experiments such as Belle II and LHCb.
\end{itemize}

In summary, we have analyzed the effects of NP coming through the neutral $Z^{\prime}$ boson on the polarized branching ratio, unpolarized
and polarized $CP$ violation asymmetries in $B_{s} \rightarrow \phi \ell^{+} \ell^{-}$ decays. We observed that
the polarized branching ratio shows a clear signal of the $Z^\prime$ model especially
for the extreme values of the parameters corresponding to this model and the values of $\mathcal{BR}_L$
and $\mathcal{BR}_T$ are almost 3 times the SM values both for $\mu$ and $\tau$ as final state leptons. It is well known that in the SM
the $CP$ violation asymmetry is negligible, where as in the present study we have seen that the unpolarized $CP$ violation asymmetry is
considerable in both $B_{s} \rightarrow \phi \mu^{+} \mu^{-}$ and $B_{s} \rightarrow \phi \tau^{+} \tau^{-}$ channels and hence it is giving a clear
message of NP arising from the neutral $Z^{\prime}$ boson.
In addition, all the polarized $CP$ violation asymmetries are significantly large
in $B_{s} \rightarrow \phi \tau^{+} \tau^{-}$ decay and they show strong dependence on the parameters of the $Z^{\prime}$ model.
Keeping in view that detection of leptons' polarization effects is really a daunting task at the experiments like the ATLAS, CMS and at LHCb, but
if we can just keep this issue aside, these $CP$ violation asymmetries which suffer less from hadronic uncertainties provide us a useful
probe to establish the NP coming through $Z^{\prime}$ model.

\section*{Acknowledgments}

The author M. J. A would like to thank the support by Quaid-i-Azam University through the University Research
Fund. M. A. P. would like to acknowledge the grant (2012/13047-2) from FAPESP. We would also like to thank R. Zwicky for recent reference about the
form factors used in the calculation.

\end{document}